\documentclass[%
reprint,
superscriptaddress,
amsmath,amssymb,
prl,
nobibnotes,
]{revtex4-2}

\usepackage[unicode=true,colorlinks=true,citecolor=blue,urlcolor=blue]{hyperref}

\usepackage{gensymb}
\usepackage{graphicx}
\usepackage{bm}

\usepackage{amsthm}
\usepackage{physics}
\usepackage{mathtools}

\allowdisplaybreaks[4]

\begin{document}

\title{Enhanced Proton Acceleration via Petawatt Laguerre-Gaussian Lasers }
\date{\today}

\affiliation{%
State Key Laboratory of High Field Laser Physics and CAS Center for Excellence in Ultra-intense Laser Science, Shanghai Institute of Optics and Fine Mechanics (SIOM), Chinese Academy of Sciences (CAS), Shanghai 201800, China
}%
\affiliation{%
University of Chinese Academy of Sciences, Beijing 100049, China}

\affiliation{%
School of Physical Science and Technology, ShanghaiTech University, Shanghai 201210, China
}%
\affiliation{%
Rutherford Appleton Laboratory, Oxfordshire OX11 0QX, United Kingdom
}%

\author{Wenpeng Wang\textsuperscript{1,2}}
\email[email: ]{wangwenpeng@siom.ac.cn}

\author{Xinyue Sun\textsuperscript{1}}

\author{Fengyu Sun\textsuperscript{1,2,3}}

\author{Zhengxing Lv\textsuperscript{1}}

\author{K. Glize\textsuperscript{4}}

\author{Zhiyong Shi\textsuperscript{1}}

\author{Yi Xu\textsuperscript{1}}

\author{Zongxin Zhang\textsuperscript{1}}

\author{Fenxiang Wu\textsuperscript{1}}

\author{Jiabing Hu\textsuperscript{1}}

\author{Jiayi Qian\textsuperscript{1}}

\author{Jiacheng Zhu\textsuperscript{1}}

\author{Xiaoyan Liang\textsuperscript{1}}

\author{Yuxin Leng\textsuperscript{1}}
\email[email: ]{lengyuxin@mail.siom.ac.cn}

\author{Ruxin Li\textsuperscript{1,2,3}}
\email[email: ]{ruxinli@siom.ac.cn}

\author{Zhizhan Xu\textsuperscript{1}}

\begin{abstract}
High-energy, high-flux collimated proton beams with high repetition rates are critical for applications such as proton therapy, proton radiography, high-energy-density matter generation, and compact particle accelerators. However, achieving proton beam collimation has typically relied on complex and expensive target fabrication or precise control of auxiliary laser pulses, which poses significant limitations for high-repetition applications. Here, we demonstrate an all-optical method for collimated proton acceleration using a single femtosecond Laguerre-Gaussian (LG) laser with an intensity exceeding $10^{20}$ W/cm$^2$ irradiating a simple planar target. Compared to conventional Gaussian laser-driven schemes, the maximum proton energy is enhanced by 60\% (reaching $\sim$ 35 MeV) and beam divergence is much reduced. Particle-in-cell simulations reveal that a plasma jet is initially focused by the hollow electric sheath field of the LG laser, then electrons in the jet are further collimated by self-generated magnetic fields. This process amplifies the charge-separation electric field between electrons and ions, leading to increased proton energy in the longitudinal direction and improved collimation in the transverse direction. This single-LG-laser-driven collimation mechanism offers a promising pathway for high-repetition, high-quality proton beam generation, with broad potential applications including proton therapy and fast ignition in inertial confinement fusion.

\end{abstract}

\maketitle
\newpage


With the advancement of ultra-intense ultrashort laser technologies, laser intensities have surpassed the previous record of $10^{22}$ W/cm$^2$ \cite{Yanovsky:08,Yoon:21}. Research on the resulting energetic ion beams driven by such intense lasers is gaining momentum owing to the distinctive characteristics of such beams, including ultrashort durations, high current densities, and low emittances \cite{RevModPhys.78.309,RevModPhys.85.751}. These beam properties can be leveraged in various applications, such as beam cancer therapy \cite{BULANOV2002240,Bulanov2002,Bulanov_2014}, proton imaging \cite{10.1063/1.1459457}, neutron production \cite{Kar_2016,PhysRevLett.110.044802}, generation of warm dense matter \cite{PhysRevLett.91.125004}, ‘‘fast ignition’’ of inertial confinement fusion targets \cite{10.1063/1.870664,PhysRevLett.102.025002}, and injectors for ion accelerators \cite{893296}. One crucial requirement for many of these applications is the reduction of beam divergence to enhance the beam brilliance and energy. However, achieving this goal proved challenging prior to 2000 because ions accelerated to energies of the order of million electron volts exhibit isotropic behavior when intense lasers are irradiated on thick solid foils \cite{10.1063/1.865510,PhysRevLett.73.1801,10.1063/1.872103}, gas jets \cite{PhysRevLett.83.737,PhysRevE.59.7042}, and submicrometric clusters \cite{Ditmire1997}. In 2000, three experimental groups reported the generation of proton beams with energies of the order of several million electron volts on the rare side of thin solid foils irradiated with high-intensity lasers \cite{PhysRevLett.84.670,PhysRevLett.84.4108,PhysRevLett.85.2945}. This mechanism, known as the target normal sheath acceleration (TNSA) \cite{10.1063/1.1333697}, is considered the most robust method for proton acceleration in experiments. Despite the low emittance and alignment along the target normal observed in the TNSA, proton beams with divergences $>10\degree$ \cite{PhysRevLett.85.2945,PhysRevLett.92.055003,PhysRevLett.90.064801} were obtained in previous experiments. As a result, these particle sources are unsuitable for ion accelerator applications. 

Various approaches have been explored to address beam collimation. One method involves constructing a target surface with shallow grooves \cite{PhysRevLett.92.204801,10.1063/1.3086424}, hemispherical shapes \cite{PhysRevLett.91.125004,10.1063/1.1333697,Ruhl2001,10.1063/1.2774001}, hemicylindrical shapes \cite{PhysRevLett.106.225003}, and cone geometries \cite{Bartal2012} to concentrate a proton beam behind the target. Additionally, auxiliary guiding cones \cite{Bartal2012,PhysRevE.87.013108,McGuffey2020}, along with rectangular and cylindrical geometries \cite{PhysRevLett.100.105004}, hollow-core microspheres \cite{Burza_2011}, and helical wires \cite{Kar2016} have been proposed to be attached to the rear of a thin target. These structures induce an electrical lens effect, thereby enhancing the beam-focusing properties. In addition to these micro-target configurations, conventional accelerator techniques involving pairs of quadrupole magnets \cite{PhysRevLett.101.055004,10.1063/1.3078291,Ter-Avetisyan_Schnürer_Polster_Nickles_Sandner_2008}, large-acceptance pulsed solenoids \cite{10.1063/1.3299363,Roth_2009}, and radio frequency electric fields \cite{PhysRevSTAB.12.063501} have been employed for energy selection or transport of laser-accelerated protons. Furthermore, proton beams can be focused using electric fields generated within a tiny cylinder \cite{doi:10.1126/science.1124412} or solenoid \cite{Ferguson_2023} by a second laser through a cascade acceleration mechanism. However, many of these methods rely on expensive and complex assembled targets \cite{PhysRevLett.91.125004,PhysRevLett.92.204801,10.1063/1.2774001,Bartal2012,PhysRevE.87.013108,McGuffey2020,PhysRevLett.100.105004,Kar2016,10.1063/1.3589476} or entail critical spatial collineation and time synchronization requirements for cascade acceleration mechanisms \cite{doi:10.1126/science.1124412,10.1063/1.5022347,10.1063/1.5088548}, thus limiting the development of high-repetition proton sources and their application in laboratory settings. 

Recently, relativistic femtosecond Laguerre-Gaussian (LG) lasers have been proposed through simulations \cite{PhysRevLett.112.215001,PhysRevLett.112.235001,Wang2015,PhysRevLett.114.173901,10.1063/1.5028555,Ju_2018,PhysRevLett.122.024801,Pae_2020,Blackman2022,Zhao2022,10.1063/5.0121973} and experimentally realized \cite{Leblanc2017,PhysRevLett.125.034801,Porat_2022}, offering a potential solution for particle collimation \cite{Wang2015,Ju_2018,Pae_2020,10.1063/5.0121973,10.1063/1.4917071,10.1063/1.4905638}. The donut-shaped laser intensity profile of an LG laser can concentrate electrons or ions along the beam axis-a unique and promising phenomenon for realizing high-repetition collimated proton acceleration. Although these ideas have been extensively researched theoretically, they have not yet been validated experimentally.

In this study, a relativistic LG laser with an intensity of $10^{20}$ W/cm$^2$ was generated using a large size reflected phase plate and used to collimate and accelerate protons at the one-petawatt (1 PW) beam line in the Shanghai Superintense Ultrafast Laser Facility (SULF). A collimated proton beam was generated using an all-optical method and verified using three-dimensional (3D) particle-in-cell (PIC) simulations. These simulations confirm the feasibility of our approach as well as elucidate the intricate electromagnetic field dynamics in the vortex laser, which drives the generation of the collimated energetic proton beam. Our experimental results provide a novel pathway for generating high-repetition tightly collimated proton beams in laboratory settings, with profound implications for general applications, including proton radiography, fast ignition of fusion targets, biomedicine, and production of warm dense matter. 

\begin{figure*}[t!]
\includegraphics[width=0.8\textwidth]{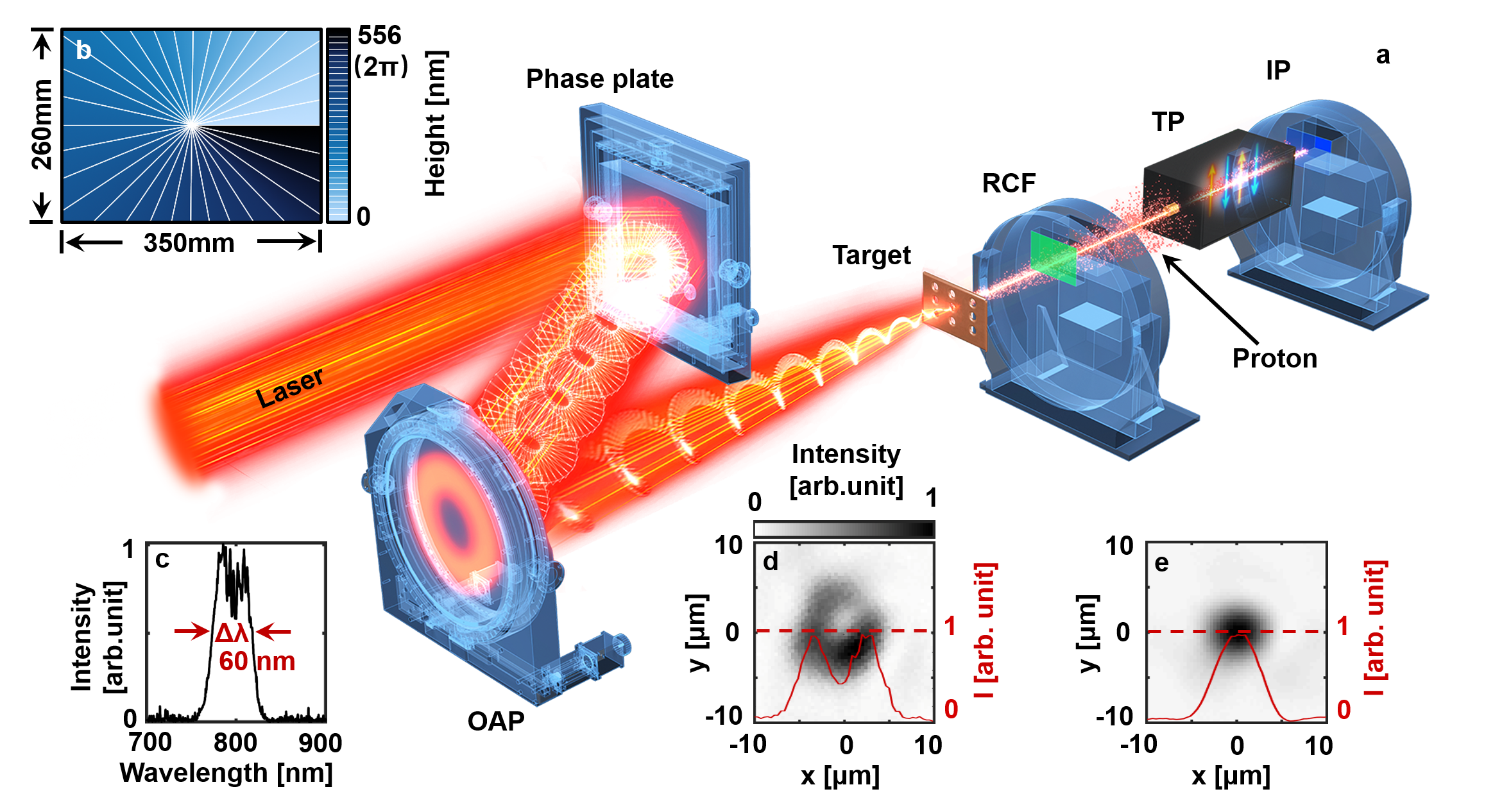}
\caption{\label{fig1} 
\textbf{Experimental setup.}
\textbf{a}, Experimental setup for generating intense LG laser and achieving proton acceleration. \textbf{b}, Design of the reflected 32-step phase plate. \textbf{c}, Measured spectrum after the final amplifier. The central wavelength of the laser is $\lambda_0=800$ nm, and the laser bandwidth is $\Delta\lambda=60$ nm. The intensity distributions of the laser focal spot of the \textbf{d}, LG and \textbf{e}, Gaussian lasers. The intensity profile across the laser center is marked by red lines (TP: Thomson parabola; IP: image plate).
}\vspace{-6pt}
\end{figure*}

\vspace{-15pt}
\subsection{Experimental Results and Discussion}\vspace{-10pt}
In the experiments, a reflected phase plate, with 32 steps at an angle of $45\degree$ relative to the direction of laser incidence, was positioned in front of an off-axis parabolic mirror (OAP) (Fig.~\ref{fig1}a). In our setup, the mixing modes are generated that are dominated by the LG$_{p=0}^{l=1}$ mode, and each step’s depth was designed to be $\sim$ 17.5 nm [$\Delta D=\lambda_0\sin45\degree/32$; here, $\lambda_0=800$ nm represents the central laser wavelength (Fig.~\ref{fig1}c)]. The dimensions of the high-reflectivity phase plate were $230\times170$ mm$^2$. A traditional Gaussian laser with a diameter of 150 mm, energy $\sim$ 15 J, and laser duration $\sim$ 28 fs is irradiated on the phase plate to generate LG mode laser based on the inner product $\left| \left< u_{lp}\left| T \right|u_{00} \right> \right|^2$ of the transfer function $T=\exp[-il(2\pi n\lambda)/(N\lambda_0)]$, where $u_{lp}$ is the arbitrary normalized LG mode amplitude at the laser waist, $\lambda$ is the laser frequency, $\lambda_0$ is the central laser frequency, and $N$ is the step number of the phase plate \cite{Longman:17}. 

The LG laser focal spot focused by an $f/4$ OAP is hollow with inner and outer radii of $\sim$ 0.6 and 4.7 $\mu$m, respectively (Fig.~\ref{fig1}c). The energy concentration of the LG laser in terms of the full width at half maximum (FWHM) of the focal spot is approximately 30\%. Hence, the laser intensity $I_\text{LG}=2.8\times10^{20}$ W/cm$^2$ can be calculated. Correspondingly, the dimensionless amplitude peak of the LG laser was $a_\text{LG}=11.4$. In contrast, the Gaussian laser produces a dot structured focal spot with a radius of $\sim$ 2.6 $\mu$m (FWHM) (Fig.~\ref{fig1}d), resulting in an intensity of $I_\text{G}=8.8\times10^{20}$ W/cm$^2$ ($a_\text{G}=20$). Both the Gaussian and LG lasers deliver the same power ($\sim$ 560 TW). The laser incidents on the 4-$\mu$m-thick Al target at an angle of $30\degree$ relative to the target normal direction in both the cases. The RCF stack was placed 8 cm behind the target to capture proton images. A Thomson parabola positioned 43 cm from the target is employed to detect the ion spectra. 

\begin{figure*}[t!]
\includegraphics[width=0.7\textwidth]{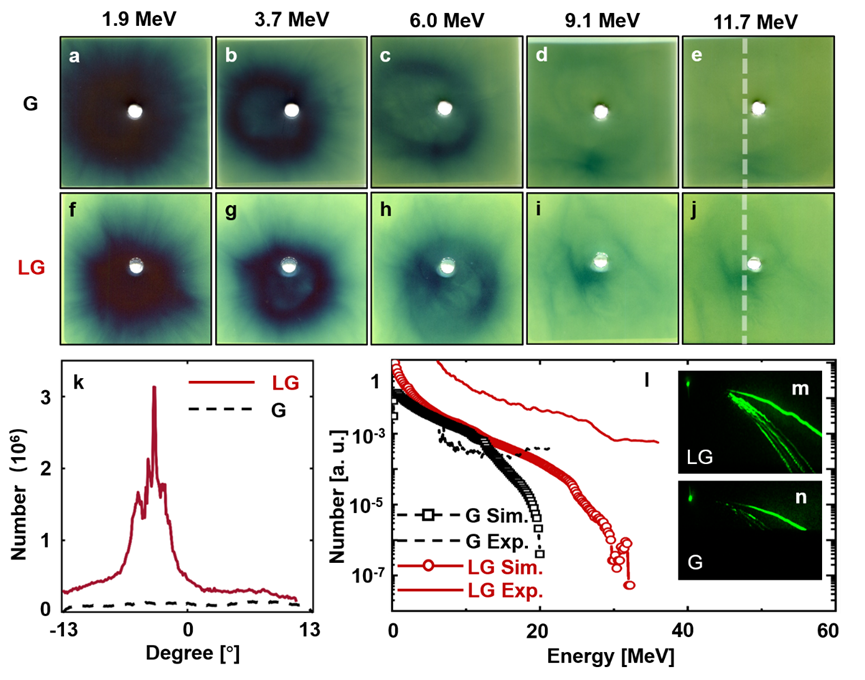}
\caption{\label{fig2} 
\textbf{Experimental result.}
Proton imaging on RCFs driven by \textbf{a-e}, a Gaussian laser and \textbf{f-j}, an LG laser in the case of 4-$\mu$m films. \textbf{k}, Divergence of the 11.7-MeV proton beam for \textbf{e}, Gaussian and \textbf{j}, LG lasers. \textbf{l}, Proton energy spectra obtained from IP plated driven by \textbf{m}, LG and \textbf{n}, G laser. 
}\vspace{-5pt}
\end{figure*}

The proton beam distributions and spectra generated by the Gaussian and LG lasers on the 4-$\mu$m Al foils are compared in Fig.~\ref{fig2}. In Figs.~\ref{fig2}a and f, the 1.9-MeV proton images on the RCF are nearly identical in both the Gaussian and LG laser cases. However, a darker dot becomes apparent for higher proton energies ($\ge6$ MeV in Figs.~\ref{fig2}h-j), indicating that the LG laser exhibits better collimation of higher-energy proton beams toward the target normal direction. The main reason is that high-energy particles can be more effectively collimated by the self-generated magnetic fields (see the simulations in Fig.\ref{fig4}m) in LG laser interactions, which are closely related to the net current of the plasma jet. In contrast, the proton beam spreads over a large area with a radius of $\sim$ 5 cm on the RCFs in the Gaussian laser case, corresponding to a divergence of $>10\degree$ in Figs.~\ref{fig2}c-e, which is typically observed in the previous TNSA experiments.

An analysis of the proton distribution is described in the \textbf{Methods} and shown in Fig.~\ref{fig2}k reveals that the proton divergence decreases to $\sim 2\degree$ for LG laser, which is much reduced compared with the Gaussian case. In addition, $\sim$ 20\% protons lie within the divergence of $2\degree$, much different from the Gaussian case, where the protons are almost uniformly distributed on the RCF (Fig.~S2). 

Notably, the maximum proton energy ($E_\text{max}$) driven by the LG lasers surpasses that of the Gaussian case, as shown in Fig.~\ref{fig2}l, where $E_\text{max}$ increases from 22 MeV to 35 MeV (a nearly 60\% increase). This means 35 MeV proton energy driven by the LG laser with an intensity of $2.8\times10^{20}$ W/cm$^2$ ($\sim$ 570 TW), which is higher than that driven by Gaussian lasers at the same laser power in our laboratory, which indicates that LG lasers may have abilities to further increase the maximum proton energy on the other PW laser facilities in the future \cite{Ziegler2024}. This enhancement in energy and reduction in divergence of the proton beams will benefit various high-precision proton applications.

Notably, all these comparisons pertain to straightforward plane targets driven by a single laser. While protons can be collimated to some degree by structured targets \cite{PhysRevLett.91.125004,PhysRevLett.92.204801,10.1063/1.2774001,Bartal2012,PhysRevE.87.013108,McGuffey2020,PhysRevLett.100.105004,Kar2016,10.1063/1.3589476} or multi laser systems \cite{doi:10.1126/science.1124412,10.1063/1.5022347,10.1063/1.5088548}, these approaches may be expensive or intricate for achieving high repetition rates with current petawatt laser systems. The intense femtosecond LG laser proposed in this study offers a practical solution for achieving collimated proton acceleration using simple plane targets and is thus promising for applications that require high-repetition beam source rates across various fields. 

\begin{figure*}[t!]
\includegraphics[width=0.6\textwidth]{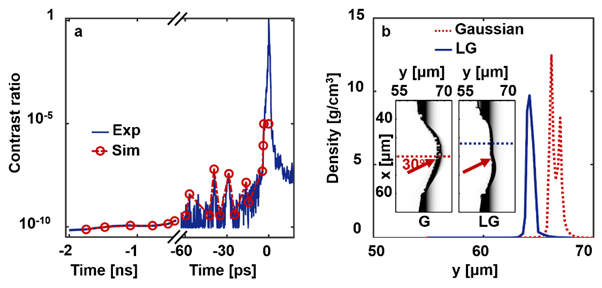}
\caption{\label{fig3} 
\textbf{a}, Profile of the ps and ns prepulse in experiments (solid line) and FLASH simulations (circle). \textbf{b}, 1D density distribution of the expanded target driven by Gaussian (dot line) and LG (solid line) prepulse in simulations. The inset plots correspond to the two-dimensional (2D) density distributions in $x-y$ plane. The dashed lines in the 2D inset plots indicate 1D density distributions.
}\vspace{-5pt}
\end{figure*}

\vspace{-15pt}
\subsection{Simulation results and discussion}\vspace{-10pt}
A distinct hollow target was generated by the hollow pre-pulse of the LG laser, which was significantly different from that of the Gaussian laser (Fig.~\ref{fig3}b). Furthermore, the front surface was more significantly displaced toward the rear of the target by the Gaussian laser than by the LG laser. This is attributed to the higher intensity of the Gaussian laser in both simulations and experiments. These expanded target distributions were then incorporated into the PIC simulations to investigate the interactions between the main pulse and plasma. 

\begin{figure*}[t!]
\includegraphics[width=0.9\textwidth]{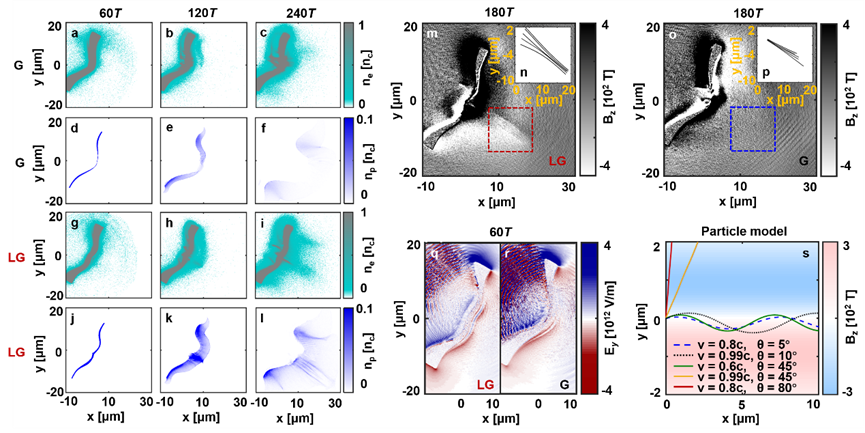}
\caption{\label{fig4} 
\textbf{PIC Simulations.}
Density distributions of \textbf{a-c}, \textbf{g-i}, electrons and \textbf{d-f}, \textbf{j-l}, proton layers driven by Gaussian lasers and LG lasers at $t=60T$, $120T$, and $240T$, respectively. Magnetic field distributions at $t=180T$ driven by \textbf{m}, LG and \textbf{o}, Gaussian laser. The proton trajectories are shown in inset plot in \textbf{n}, and \textbf{p}. Electric field distributions are plot at $t =60T$ for \textbf{q}, LG and \textbf{r}, Gaussian lasers. \textbf{s}, Electron trajectories in the magnetic field in the calculation.
}\vspace{-5pt}
\end{figure*}

Figure~\ref{fig4} shows the proton-collimated acceleration induced by the LG lasers. Upon laser interaction with the front surface of the target at $t\sim30T$, a fraction of electrons accelerates toward the rear of the target (Figs.~\ref{fig4}a and g), establishing a charge-separation electric field between the hot electrons and ions (Figs.~\ref{fig4}q, r). This process leads to the ionization of vapor or pollutants at the rear surface, generating a proton layer via the sheath electric field. These protons experience further acceleration in the direction normal to the target, which is known as the TNSA mechanism. 

For LG lasers, the back surface of the target is initially curved inward due to the prepulse of the LG laser, which leads to the inward acceleration of the electron beam at the onset of the TNSA process (Fig.~\ref{fig4}h). This curvature enhances the charge separation field between the electrons and ions and directs it along the axis, which may contribute to higher proton acceleration and reduced beam divergence in the initial TNSA stage, as shown in Figs.~\ref{fig4}h and k. Notably, particle dispersion, which typically occurs beyond the focusing position in the absence of additional confining forces, is averted because of the significant predominance of electrons over protons in the jet. This results in a negative current that generates a magnetic tunnel around the jet, confining the electrons within it (Fig.~\ref{fig4}m). Here, electrons at an angle $\theta$ relative to the axial ($x$) direction can bend inward and be confined within a magnetic field tunnel if the collimation condition $r_\text{B}\ge r_\text{e}(1-\cos\theta)$ is satisfied. $r_\text{e}=\gamma_\text{e} m_\text{e}v_\text{e}/(qB)$ is the electron Larmor radius, $\gamma_\text{e}= (1-v_\text{e}^2/c^2)^{-1/2}$ is the relativistic factor, $v_\text{e}$ is the electron velocity, $m_\text{e}$ is the electron mass, and $q$ is the electron charge. For instance, electrons with $v_\text{e}=0.8c$ and $\theta=5\degree$ are collimated within the plasma jet under such a collimation condition (Fig.~\ref{fig4}s). In this way, the confined electron within the plasma jet maintains a higher charge separation field, which, in turn, more efficiently accelerates protons during the subsequent stages of TNSA. Ultimately, both collimation and the maximum energy of the proton beam can be further enhanced in our LG laser case (Fig.~\ref{fig4}n), which is significantly different from the dispersion movement of protons driven by Gaussian lasers, as shown in Fig.~\ref{fig4}p.

In addition, protons are dispersed at a wider angle in Figs.~\ref{fig4}d-f, which is attributed to the Gaussian-like plasma target driven by traditional Gaussian lasers. This reduces the strength of the sheath field in the Gaussian laser case (Fig.~\ref{fig4}r), producing a lower maximum energy of the proton beam than that in the LG laser case, which is consistent with the experimental results (Fig.~\ref{fig2}l). Previously, an ultrahigh-contrast laser was proposed to maintain the flat back surface of the target, thus mitigating the beam divergence and increasing the beam maximum energy to some extent. However, further increasing the maximum energy of the proton beam requires an ultrathin target, and laser contrast remains problematic. Moreover, the issue of prepulse has become exacerbated with advancements in laser energy, particularly with the recent emergence of 10 PW laser settings. Despite efforts to manage the contrast, the pre-pulse continued to degrade the back surface of the target. Fortunately, this challenge can be overcome by employing intense LG lasers, where the hollow laser intensity generates a focusing sheath field, enhancing the TNSA mechanism. 

	\vspace{-15pt}
	\subsection{Conclusion}\vspace{-10pt}
In conclusion, the experimental results demonstrated successful realization of collimated proton acceleration driven by LG lasers. Compared with traditional Gaussian lasers, this approach yields a 60\% increase in the maximum proton energy and the beam divergence is much reduced. PIC simulations and theoretical calculations indicate that the hollow sheath field initially concentrates the plasma with a negative current at the back surface of the target, which is further collimated by self-generated magnetic fields. This enhancement of the traditional TNSA is promising for various applications, such as proton therapy, imaging, and fast ignition in the initial confined fusion. However, although this study marks the beginning of inaugural experiments on collimated particle beams propelled by LG lasers, there remains room for improvement. For instance, continuous phase plates, which can refine the hollow focal spot of LG lasers, thus producing highly collimated accelerated beams, can be employed in future experiments.

\vspace{-15pt}
\subsection{Methods}\vspace{-10pt}\label{Methods}
\textbf{FLASH simulation.}
To account for prepulse effects in the experiments, we used the hydrodynamic code FLASH \cite{Fryxell_2000} to simulate the expansion of the Al target following interactions with picosecond (ps) and nanosecond (ns) prepulses (Fig.~\ref{fig3}). The results were then used as initial conditions for particle-in-cell (PIC) simulations. FLASH employs the Euler equations to solve hydrodynamic equations by integrating laser energy deposition, radiation transport, and electron thermal conduction. The laser ray-tracing method, along with the inverse-bremsstrahlung deposition mechanism, was applied in the simulation \cite{Wang_Jiang_Li_Dong_Shen_Leng_Li_Xu_2019}. For electron thermal conduction, the classical Spitzer–Harm model is utilized \cite{PhysRev.89.977}. The program approximates the radiative transfer equation and the electron energy equation using the multigroup diffusion approximation. Coefficients not provided by FLASH are generated in tabular form using the MPQEOS code \cite{KEMP1998674}, while mean opacity coefficients are computed using the SNOP code \cite{Eidmann_1994}.

The size of the FLASH simulation area was $60(x)\times60(y)$ $\mu\text{m}^2$ and consisting of $1000\times1000$ cells. Initially, the front surface of the target is positioned at $y=60\ \mu$m (parallel to the $x$ axis) with a thickness of 4 $\mu$m and an initial density of 2.7 g/cm$^3$. The beam cross-sectional function initially characterizes the power distribution of the rays within the LG laser beam. The transverse intensity distribution of Gaussian laser focal spot is described by a weighting function, $\omega=\exp[-(r/R_r)^{2\gamma}]$, where $0<r<8$ $\mu$m is the radius region of the focal spot, $R_x=3.3$ $\mu$m is the decay radius, and $\gamma=1.1$ is the Gaussian super exponent. Consequently, the diameter of the Gaussian focal spot is 5.2 $\mu$m (FWHM), like the experimental setup in Fig.~\ref{fig1}e. Subsequently, two Gaussian expressions are combined to depict the energy distribution of the LG laser. Each Gaussian focal spot has a radius of 5 $\mu$m, and the separation between the centers of these two Gaussian shapes is 6 $\mu$m (Fig.~\ref{fig1}d). The 2-ns prepulse is incident on the Al target at a $30\degree$ angle relative to the target normal direction (Fig.~\ref{fig3}b). 

\textbf{PIC simulation.}
In the three-dimensional (3D) PIC simulations (EPOCH code \cite{Arber_2015}), the laser propagated from the left boundary of the simulation box in the $+x$ direction, irradiating the target at an angle of $30\degree$. The linearly polarized LG laser amplitude $a_\text{LG}=11.4$ and Gaussian laser amplitude ($a_\text{G}=20$) were set in the simulation to match those used in the experimental setup shown in Fig.~\ref{fig1}. Both the lasers had a duration of 28 fs. The amplitudes of the LG and Gaussian lasers can be expressed as: 
\begin{equation}
	\begin{aligned}
		a_y\left( \text{LG}_{p}^{l} \right) &=a_{\text{LG}}\left( -1 \right) ^pX^{1/2}L_{p}^{l}\left( X \right) \exp \left( -X/2 \right) g\left( x-ct \right) 
		\\
		&\times \cos \left( kx-\omega t+l\varphi +\theta \right),
	\end{aligned}
\end{equation}
\begin{equation}
	a_y\left( \text{G} \right) =a_{\text{G}}\exp \left( -X/2 \right) g\left( x-ct \right) \cos \left( kx-\omega t+\theta \right) ,
\end{equation}
where the laser is polarized along $y-$axis and $p=0$ and $l=1$ are the radial mode number and topological charge, respectively. The variable $X=(\sqrt{2}r/w_\bot)^2$, where $r$ is the radial component in cylindrical coordinates, the beam waist sizes for LG and Gaussian laser are $w_{\text{LG},\bot}=3$ $\mu$m and $w_{\text{G},\bot}=2.5$ $\mu$m, respectively. $L_p^l(X)$ is a generalized Laguerre polynomial. The time profile $g(x-ct)=\sin^2[\pi(x-ct)/T]$ is defined within a pulse duration $T=28$ fs in the $x$ direction. $\varphi$ is azimuth angle and $\theta$ is laser initial phase, $\omega$ is the laser frequency, $k=2\pi/\lambda$ is the wave number, and the laser wavelength is $\lambda=0.8$ $\mu$m.

The expanded Al target is positioned obliquely at $x=0$ $\mu$m and is coated with a proton layer 0.3 $\mu$m thick, where the proton density equal to the electron density ($n_\text{p}=n_\text{e}=1\ n_\text{c}$). To be simple, we input the half distribution of the FLASH targets with the center axis ($x=50$ $\mu$m in Fig.~\ref{fig3}b) into the 3D simulations, which is symmetrically rotated around the target normal direction to form a Gaussian and hollow plasma in Fig.~\ref{fig4}. For computational efficiency, the maximum electron density in the Al layer was set to $n_\text{e}=35n_\text{c}$, whereas the density distributions for the other components ($n_\text{e}<35n_\text{c}$) in the longitudinal direction retained the characteristics shown in Fig.~\ref{fig3}b. Here, the Al foil was assumed to be averagely ionized into Al$^{7+}$ ions. The proton layer extended along the density profile at the back of the target in the case of the Gaussian laser, as shown in Fig.~\ref{fig3}b. The size of the simulation box was $100(x)\times60(y)\times60(z)$ $\mu$m$^3$, with a grid of $2000\times1200\times1200$ cells, each filled with five electrons and five ions, for accurate representation and computational efficiency. 

\textbf{Analysis of RCFs.}
We describe the analyzing progress for the data on RCFs. Firstly, one-dimensional spectrum deconvolution approach is introduced. Here, we ignore the spatial distribution of particles in three dimensions and consider only the distribution of particle numbers in the $z-$direction, where the $z-$direction corresponds to ion energy ($E$). Essentially, we are solving for the relationship between $\text{d}N/\text{d}E$ and $E$, which has been researched by Schollmeier \textit{et al}. \cite{10.1063/1.4870895}. 

Secondly, the processing of dose information is introduced. The color information is read from the red channel of a 16-bit image. After removing the RCF background, the color information is converted to OD values. These OD values are then converted to dose information using calibration data from the literature (Fig.~S1). Since high resolution is not required, we combine $10\times10$ pixels into one pixel. The processing procedure is illustrated in the following example figures.

For processing the high-intensity region of the collimation beam, two methods are employed: one method involves identifying the maximum value point and selecting the envelope of the region where the values are a certain proportion of the peak value; the other method involves selecting a circle of a determined size and position that includes the high-intensity region. The former is used to analyze dose concentration, while the latter is used for spectrum deconvolution. 

Thirdly, the processing of the response function on RCF is introduced. For the response function $R(E)$, we use SRIM for simulation. SRIM is a computational software that simulates the interaction of ions with matter. It uses the Monte Carlo simulation method to track the motion of incident particles, yielding the expected values of various physical quantities and their corresponding statistical errors. Here, we use the PySrim data library in Python to uniformly sample incident ion energies in the range of 1.0-15.0 MeV at intervals of 0.02 MeV. This results in the response functions for 14 layers of RCF, as shown in the following figure.

Fourthly, processing of exponential form function spectrum assumption is introduced. Assuming the spectrum is in exponential form $f(E)=\text{d}N/\text{d}E$, that is $f(E)=\text{d}N/\text{d}E=N_0/E\exp [-E/(k_{\text{B}}T)] $. $N_0$ and $\beta=1/(k_\text{B}T)$ are unknown. Based on the data on RCF we solved previously, the Newton least squares method is used for iterative solving, with the maximum number of iterations set to 30. 

Finally, the proton spectra from RCFs are presented in Fig.~S2.


\begin{thebibliography}{73}%
	\makeatletter
	\providecommand \@ifxundefined [1]{%
		\@ifx{#1\undefined}
	}%
	\providecommand \@ifnum [1]{%
		\ifnum #1\expandafter \@firstoftwo
		\else \expandafter \@secondoftwo
		\fi
	}%
	\providecommand \@ifx [1]{%
		\ifx #1\expandafter \@firstoftwo
		\else \expandafter \@secondoftwo
		\fi
	}%
	\providecommand \natexlab [1]{#1}%
	\providecommand \enquote  [1]{``#1''}%
	\providecommand \bibnamefont  [1]{#1}%
	\providecommand \bibfnamefont [1]{#1}%
	\providecommand \citenamefont [1]{#1}%
	\providecommand \href@noop [0]{\@secondoftwo}%
	\providecommand \href [0]{\begingroup \@sanitize@url \@href}%
	\providecommand \@href[1]{\@@startlink{#1}\@@href}%
	\providecommand \@@href[1]{\endgroup#1\@@endlink}%
	\providecommand \@sanitize@url [0]{\catcode `\\12\catcode `\$12\catcode
		`\&12\catcode `\#12\catcode `\^12\catcode `\_12\catcode `\%12\relax}%
	\providecommand \@@startlink[1]{}%
	\providecommand \@@endlink[0]{}%
	\providecommand \url  [0]{\begingroup\@sanitize@url \@url }%
	\providecommand \@url [1]{\endgroup\@href {#1}{\urlprefix }}%
	\providecommand \urlprefix  [0]{URL }%
	\providecommand \Eprint [0]{\href }%
	\providecommand \doibase [0]{https://doi.org/}%
	\providecommand \selectlanguage [0]{\@gobble}%
	\providecommand \bibinfo  [0]{\@secondoftwo}%
	\providecommand \bibfield  [0]{\@secondoftwo}%
	\providecommand \translation [1]{[#1]}%
	\providecommand \BibitemOpen [0]{}%
	\providecommand \bibitemStop [0]{}%
	\providecommand \bibitemNoStop [0]{.\EOS\space}%
	\providecommand \EOS [0]{\spacefactor3000\relax}%
	\providecommand \BibitemShut  [1]{\csname bibitem#1\endcsname}%
	\let\auto@bib@innerbib\@empty
	\bibitem [{\citenamefont {Yanovsky}\ \emph {et~al.}(2008)\citenamefont
		{Yanovsky}, \citenamefont {Chvykov}, \citenamefont {Kalinchenko},
		\citenamefont {Rousseau}, \citenamefont {Planchon}, \citenamefont {Matsuoka},
		\citenamefont {Maksimchuk}, \citenamefont {Nees}, \citenamefont {Cheriaux},
		\citenamefont {Mourou},\ and\ \citenamefont {Krushelnick}}]{Yanovsky:08}%
	\BibitemOpen
	\bibfield  {author} {\bibinfo {author} {\bibfnamefont {V.}~\bibnamefont
			{Yanovsky}}, \bibinfo {author} {\bibfnamefont {V.}~\bibnamefont {Chvykov}},
		\bibinfo {author} {\bibfnamefont {G.}~\bibnamefont {Kalinchenko}}, \bibinfo
		{author} {\bibfnamefont {P.}~\bibnamefont {Rousseau}}, \bibinfo {author}
		{\bibfnamefont {T.}~\bibnamefont {Planchon}}, \bibinfo {author}
		{\bibfnamefont {T.}~\bibnamefont {Matsuoka}}, \bibinfo {author}
		{\bibfnamefont {A.}~\bibnamefont {Maksimchuk}}, \bibinfo {author}
		{\bibfnamefont {J.}~\bibnamefont {Nees}}, \bibinfo {author} {\bibfnamefont
			{G.}~\bibnamefont {Cheriaux}}, \bibinfo {author} {\bibfnamefont
			{G.}~\bibnamefont {Mourou}},\ and\ \bibinfo {author} {\bibfnamefont
			{K.}~\bibnamefont {Krushelnick}},\ }\bibfield  {title} {\bibinfo {title}
		{Ultra-high intensity-300-\text{TW} laser at 0.1 \text{Hz} repetition
			rate.},\ }\href {https://doi.org/10.1364/OE.16.002109} {\bibfield  {journal}
		{\bibinfo  {journal} {Opt. Express}\ }\textbf {\bibinfo {volume} {16}},\
		\bibinfo {pages} {2109} (\bibinfo {year} {2008})}\BibitemShut {NoStop}%
	\bibitem [{\citenamefont {Yoon}\ \emph {et~al.}(2021)\citenamefont {Yoon},
		\citenamefont {Kim}, \citenamefont {Choi}, \citenamefont {Sung},
		\citenamefont {Lee}, \citenamefont {Lee},\ and\ \citenamefont
		{Nam}}]{Yoon:21}%
	\BibitemOpen
	\bibfield  {author} {\bibinfo {author} {\bibfnamefont {J.~W.}\ \bibnamefont
			{Yoon}}, \bibinfo {author} {\bibfnamefont {Y.~G.}\ \bibnamefont {Kim}},
		\bibinfo {author} {\bibfnamefont {I.~W.}\ \bibnamefont {Choi}}, \bibinfo
		{author} {\bibfnamefont {J.~H.}\ \bibnamefont {Sung}}, \bibinfo {author}
		{\bibfnamefont {H.~W.}\ \bibnamefont {Lee}}, \bibinfo {author} {\bibfnamefont
			{S.~K.}\ \bibnamefont {Lee}},\ and\ \bibinfo {author} {\bibfnamefont {C.~H.}\
			\bibnamefont {Nam}},\ }\bibfield  {title} {\bibinfo {title} {Realization of
			laser intensity over $10^{23}$ \text{W}/cm$^2$},\ }\href
	{https://doi.org/10.1364/OPTICA.420520} {\bibfield  {journal} {\bibinfo
			{journal} {Optica}\ }\textbf {\bibinfo {volume} {8}},\ \bibinfo {pages} {630}
		(\bibinfo {year} {2021})}\BibitemShut {NoStop}%
	\bibitem [{\citenamefont {Mourou}\ \emph {et~al.}(2006)\citenamefont {Mourou},
		\citenamefont {Tajima},\ and\ \citenamefont {Bulanov}}]{RevModPhys.78.309}%
	\BibitemOpen
	\bibfield  {author} {\bibinfo {author} {\bibfnamefont {G.~A.}\ \bibnamefont
			{Mourou}}, \bibinfo {author} {\bibfnamefont {T.}~\bibnamefont {Tajima}},\
		and\ \bibinfo {author} {\bibfnamefont {S.~V.}\ \bibnamefont {Bulanov}},\
	}\bibfield  {title} {\bibinfo {title} {Optics in the relativistic regime},\
	}\href {https://doi.org/10.1103/RevModPhys.78.309} {\bibfield  {journal}
		{\bibinfo  {journal} {Rev. Mod. Phys.}\ }\textbf {\bibinfo {volume} {78}},\
		\bibinfo {pages} {309} (\bibinfo {year} {2006})}\BibitemShut {NoStop}%
	\bibitem [{\citenamefont {Macchi}\ \emph {et~al.}(2013)\citenamefont {Macchi},
		\citenamefont {Borghesi},\ and\ \citenamefont {Passoni}}]{RevModPhys.85.751}%
	\BibitemOpen
	\bibfield  {author} {\bibinfo {author} {\bibfnamefont {A.}~\bibnamefont
			{Macchi}}, \bibinfo {author} {\bibfnamefont {M.}~\bibnamefont {Borghesi}},\
		and\ \bibinfo {author} {\bibfnamefont {M.}~\bibnamefont {Passoni}},\
	}\bibfield  {title} {\bibinfo {title} {Ion acceleration by superintense
			laser-plasma interaction},\ }\href
	{https://doi.org/10.1103/RevModPhys.85.751} {\bibfield  {journal} {\bibinfo
			{journal} {Rev. Mod. Phys.}\ }\textbf {\bibinfo {volume} {85}},\ \bibinfo
		{pages} {751} (\bibinfo {year} {2013})}\BibitemShut {NoStop}%
	\bibitem [{\citenamefont {Bulanov}\ \emph {et~al.}(2002)\citenamefont
		{Bulanov}, \citenamefont {Esirkepov}, \citenamefont {Khoroshkov},
		\citenamefont {Kuznetsov},\ and\ \citenamefont {Pegoraro}}]{BULANOV2002240}%
	\BibitemOpen
	\bibfield  {author} {\bibinfo {author} {\bibfnamefont {S.}~\bibnamefont
			{Bulanov}}, \bibinfo {author} {\bibfnamefont {T.}~\bibnamefont {Esirkepov}},
		\bibinfo {author} {\bibfnamefont {V.}~\bibnamefont {Khoroshkov}}, \bibinfo
		{author} {\bibfnamefont {A.}~\bibnamefont {Kuznetsov}},\ and\ \bibinfo
		{author} {\bibfnamefont {F.}~\bibnamefont {Pegoraro}},\ }\bibfield  {title}
	{\bibinfo {title} {Oncological hadrontherapy with laser ion accelerators},\
	}\href {https://doi.org/https://doi.org/10.1016/S0375-9601(02)00521-2}
	{\bibfield  {journal} {\bibinfo  {journal} {Phys. Lett. A}\ }\textbf
		{\bibinfo {volume} {299}},\ \bibinfo {pages} {240} (\bibinfo {year}
		{2002})}\BibitemShut {NoStop}%
	\bibitem [{\citenamefont {Bulanov}\ and\ \citenamefont
		{Khoroshkov}(2002)}]{Bulanov2002}%
	\BibitemOpen
	\bibfield  {author} {\bibinfo {author} {\bibfnamefont {S.~V.}\ \bibnamefont
			{Bulanov}}\ and\ \bibinfo {author} {\bibfnamefont {V.~S.}\ \bibnamefont
			{Khoroshkov}},\ }\bibfield  {title} {\bibinfo {title} {Feasibility of using
			laser ion accelerators in proton therapy},\ }\href
	{https://doi.org/10.1134/1.1478534} {\bibfield  {journal} {\bibinfo
			{journal} {Plasma Phys. Rep.}\ }\textbf {\bibinfo {volume} {28}},\ \bibinfo
		{pages} {453} (\bibinfo {year} {2002})}\BibitemShut {NoStop}%
	\bibitem [{\citenamefont {Bulanov}\ \emph {et~al.}(2014)\citenamefont
		{Bulanov}, \citenamefont {Wilkens}, \citenamefont {Esirkepov}, \citenamefont
		{Korn}, \citenamefont {Kraft}, \citenamefont {Kraft}, \citenamefont {Molls},\
		and\ \citenamefont {Khoroshkov}}]{Bulanov_2014}%
	\BibitemOpen
	\bibfield  {author} {\bibinfo {author} {\bibfnamefont {S.~V.}\ \bibnamefont
			{Bulanov}}, \bibinfo {author} {\bibfnamefont {J.~J.}\ \bibnamefont
			{Wilkens}}, \bibinfo {author} {\bibfnamefont {T.~Z.}\ \bibnamefont
			{Esirkepov}}, \bibinfo {author} {\bibfnamefont {G.}~\bibnamefont {Korn}},
		\bibinfo {author} {\bibfnamefont {G.}~\bibnamefont {Kraft}}, \bibinfo
		{author} {\bibfnamefont {S.~D.}\ \bibnamefont {Kraft}}, \bibinfo {author}
		{\bibfnamefont {M.}~\bibnamefont {Molls}},\ and\ \bibinfo {author}
		{\bibfnamefont {V.~S.}\ \bibnamefont {Khoroshkov}},\ }\bibfield  {title}
	{\bibinfo {title} {Laser ion acceleration for hadron therapy},\ }\href
	{https://doi.org/10.3367/UFNe.0184.201412a.1265} {\bibfield  {journal}
		{\bibinfo  {journal} {Physics-Uspekhi}\ }\textbf {\bibinfo {volume} {57}},\
		\bibinfo {pages} {1149} (\bibinfo {year} {2014})}\BibitemShut {NoStop}%
	\bibitem [{\citenamefont {Borghesi}\ \emph {et~al.}(2002)\citenamefont
		{Borghesi}, \citenamefont {Campbell}, \citenamefont {Schiavi}, \citenamefont
		{Haines}, \citenamefont {Willi}, \citenamefont {MacKinnon}, \citenamefont
		{Patel}, \citenamefont {Gizzi}, \citenamefont {Galimberti}, \citenamefont
		{Clarke}, \citenamefont {Pegoraro}, \citenamefont {Ruhl},\ and\ \citenamefont
		{Bulanov}}]{10.1063/1.1459457}%
	\BibitemOpen
	\bibfield  {author} {\bibinfo {author} {\bibfnamefont {M.}~\bibnamefont
			{Borghesi}}, \bibinfo {author} {\bibfnamefont {D.~H.}\ \bibnamefont
			{Campbell}}, \bibinfo {author} {\bibfnamefont {A.}~\bibnamefont {Schiavi}},
		\bibinfo {author} {\bibfnamefont {M.~G.}\ \bibnamefont {Haines}}, \bibinfo
		{author} {\bibfnamefont {O.}~\bibnamefont {Willi}}, \bibinfo {author}
		{\bibfnamefont {A.~J.}\ \bibnamefont {MacKinnon}}, \bibinfo {author}
		{\bibfnamefont {P.}~\bibnamefont {Patel}}, \bibinfo {author} {\bibfnamefont
			{L.~A.}\ \bibnamefont {Gizzi}}, \bibinfo {author} {\bibfnamefont
			{M.}~\bibnamefont {Galimberti}}, \bibinfo {author} {\bibfnamefont {R.~J.}\
			\bibnamefont {Clarke}}, \bibinfo {author} {\bibfnamefont {F.}~\bibnamefont
			{Pegoraro}}, \bibinfo {author} {\bibfnamefont {H.}~\bibnamefont {Ruhl}},\
		and\ \bibinfo {author} {\bibfnamefont {S.}~\bibnamefont {Bulanov}},\
	}\bibfield  {title} {\bibinfo {title} {Electric field detection in
			laser-plasma interaction experiments via the proton imaging technique},\
	}\href {https://doi.org/10.1063/1.1459457} {\bibfield  {journal} {\bibinfo
			{journal} {Phys. Plasmas}\ }\textbf {\bibinfo {volume} {9}},\ \bibinfo
		{pages} {2214} (\bibinfo {year} {2002})}\BibitemShut {NoStop}%
	\bibitem [{\citenamefont {Kar}\ \emph {et~al.}(2016{\natexlab{a}})\citenamefont
		{Kar}, \citenamefont {Green}, \citenamefont {Ahmed}, \citenamefont {Alejo},
		\citenamefont {Robinson}, \citenamefont {Cerchez}, \citenamefont {Clarke},
		\citenamefont {Doria}, \citenamefont {Dorkings}, \citenamefont {Fernandez},
		\citenamefont {Mirfayzi}, \citenamefont {McKenna}, \citenamefont {Naughton},
		\citenamefont {Neely}, \citenamefont {Norreys}, \citenamefont {Peth},
		\citenamefont {Powell}, \citenamefont {Ruiz}, \citenamefont {Swain},
		\citenamefont {Willi},\ and\ \citenamefont {Borghesi}}]{Kar_2016}%
	\BibitemOpen
	\bibfield  {author} {\bibinfo {author} {\bibfnamefont {S.}~\bibnamefont
			{Kar}}, \bibinfo {author} {\bibfnamefont {A.}~\bibnamefont {Green}}, \bibinfo
		{author} {\bibfnamefont {H.}~\bibnamefont {Ahmed}}, \bibinfo {author}
		{\bibfnamefont {A.}~\bibnamefont {Alejo}}, \bibinfo {author} {\bibfnamefont
			{A.~P.~L.}\ \bibnamefont {Robinson}}, \bibinfo {author} {\bibfnamefont
			{M.}~\bibnamefont {Cerchez}}, \bibinfo {author} {\bibfnamefont
			{R.}~\bibnamefont {Clarke}}, \bibinfo {author} {\bibfnamefont
			{D.}~\bibnamefont {Doria}}, \bibinfo {author} {\bibfnamefont
			{S.}~\bibnamefont {Dorkings}}, \bibinfo {author} {\bibfnamefont
			{J.}~\bibnamefont {Fernandez}}, \bibinfo {author} {\bibfnamefont {S.~R.}\
			\bibnamefont {Mirfayzi}}, \bibinfo {author} {\bibfnamefont {P.}~\bibnamefont
			{McKenna}}, \bibinfo {author} {\bibfnamefont {K.}~\bibnamefont {Naughton}},
		\bibinfo {author} {\bibfnamefont {D.}~\bibnamefont {Neely}}, \bibinfo
		{author} {\bibfnamefont {P.}~\bibnamefont {Norreys}}, \bibinfo {author}
		{\bibfnamefont {C.}~\bibnamefont {Peth}}, \bibinfo {author} {\bibfnamefont
			{H.}~\bibnamefont {Powell}}, \bibinfo {author} {\bibfnamefont {J.~A.}\
			\bibnamefont {Ruiz}}, \bibinfo {author} {\bibfnamefont {J.}~\bibnamefont
			{Swain}}, \bibinfo {author} {\bibfnamefont {O.}~\bibnamefont {Willi}},\ and\
		\bibinfo {author} {\bibfnamefont {M.}~\bibnamefont {Borghesi}},\ }\bibfield
	{title} {\bibinfo {title} {Beamed neutron emission driven by laser
			accelerated light ions},\ }\href
	{https://doi.org/10.1088/1367-2630/18/5/053002} {\bibfield  {journal}
		{\bibinfo  {journal} {New J. Phys.}\ }\textbf {\bibinfo {volume} {18}},\
		\bibinfo {pages} {053002} (\bibinfo {year} {2016}{\natexlab{a}})}\BibitemShut
	{NoStop}%
	\bibitem [{\citenamefont {Roth}\ \emph {et~al.}(2013)\citenamefont {Roth},
		\citenamefont {Jung}, \citenamefont {Falk}, \citenamefont {Guler},
		\citenamefont {Deppert}, \citenamefont {Devlin}, \citenamefont {Favalli},
		\citenamefont {Fernandez}, \citenamefont {Gautier}, \citenamefont {Geissel},
		\citenamefont {Haight}, \citenamefont {Hamilton}, \citenamefont {Hegelich},
		\citenamefont {Johnson}, \citenamefont {Merrill}, \citenamefont {Schaumann},
		\citenamefont {Schoenberg}, \citenamefont {Schollmeier}, \citenamefont
		{Shimada}, \citenamefont {Taddeucci}, \citenamefont {Tybo}, \citenamefont
		{Wagner}, \citenamefont {Wender}, \citenamefont {Wilde},\ and\ \citenamefont
		{Wurden}}]{PhysRevLett.110.044802}%
	\BibitemOpen
	\bibfield  {author} {\bibinfo {author} {\bibfnamefont {M.}~\bibnamefont
			{Roth}}, \bibinfo {author} {\bibfnamefont {D.}~\bibnamefont {Jung}}, \bibinfo
		{author} {\bibfnamefont {K.}~\bibnamefont {Falk}}, \bibinfo {author}
		{\bibfnamefont {N.}~\bibnamefont {Guler}}, \bibinfo {author} {\bibfnamefont
			{O.}~\bibnamefont {Deppert}}, \bibinfo {author} {\bibfnamefont
			{M.}~\bibnamefont {Devlin}}, \bibinfo {author} {\bibfnamefont
			{A.}~\bibnamefont {Favalli}}, \bibinfo {author} {\bibfnamefont
			{J.}~\bibnamefont {Fernandez}}, \bibinfo {author} {\bibfnamefont
			{D.}~\bibnamefont {Gautier}}, \bibinfo {author} {\bibfnamefont
			{M.}~\bibnamefont {Geissel}}, \bibinfo {author} {\bibfnamefont
			{R.}~\bibnamefont {Haight}}, \bibinfo {author} {\bibfnamefont {C.~E.}\
			\bibnamefont {Hamilton}}, \bibinfo {author} {\bibfnamefont {B.~M.}\
			\bibnamefont {Hegelich}}, \bibinfo {author} {\bibfnamefont {R.~P.}\
			\bibnamefont {Johnson}}, \bibinfo {author} {\bibfnamefont {F.}~\bibnamefont
			{Merrill}}, \bibinfo {author} {\bibfnamefont {G.}~\bibnamefont {Schaumann}},
		\bibinfo {author} {\bibfnamefont {K.}~\bibnamefont {Schoenberg}}, \bibinfo
		{author} {\bibfnamefont {M.}~\bibnamefont {Schollmeier}}, \bibinfo {author}
		{\bibfnamefont {T.}~\bibnamefont {Shimada}}, \bibinfo {author} {\bibfnamefont
			{T.}~\bibnamefont {Taddeucci}}, \bibinfo {author} {\bibfnamefont {J.~L.}\
			\bibnamefont {Tybo}}, \bibinfo {author} {\bibfnamefont {F.}~\bibnamefont
			{Wagner}}, \bibinfo {author} {\bibfnamefont {S.~A.}\ \bibnamefont {Wender}},
		\bibinfo {author} {\bibfnamefont {C.~H.}\ \bibnamefont {Wilde}},\ and\
		\bibinfo {author} {\bibfnamefont {G.~A.}\ \bibnamefont {Wurden}},\ }\bibfield
	{title} {\bibinfo {title} {Bright laser-driven neutron source based on the
			relativistic transparency of solids},\ }\href
	{https://doi.org/10.1103/PhysRevLett.110.044802} {\bibfield  {journal}
		{\bibinfo  {journal} {Phys. Rev. Lett.}\ }\textbf {\bibinfo {volume} {110}},\
		\bibinfo {pages} {044802} (\bibinfo {year} {2013})}\BibitemShut {NoStop}%
	\bibitem [{\citenamefont {Patel}\ \emph {et~al.}(2003)\citenamefont {Patel},
		\citenamefont {Mackinnon}, \citenamefont {Key}, \citenamefont {Cowan},
		\citenamefont {Foord}, \citenamefont {Allen}, \citenamefont {Price},
		\citenamefont {Ruhl}, \citenamefont {Springer},\ and\ \citenamefont
		{Stephens}}]{PhysRevLett.91.125004}%
	\BibitemOpen
	\bibfield  {author} {\bibinfo {author} {\bibfnamefont {P.~K.}\ \bibnamefont
			{Patel}}, \bibinfo {author} {\bibfnamefont {A.~J.}\ \bibnamefont
			{Mackinnon}}, \bibinfo {author} {\bibfnamefont {M.~H.}\ \bibnamefont {Key}},
		\bibinfo {author} {\bibfnamefont {T.~E.}\ \bibnamefont {Cowan}}, \bibinfo
		{author} {\bibfnamefont {M.~E.}\ \bibnamefont {Foord}}, \bibinfo {author}
		{\bibfnamefont {M.}~\bibnamefont {Allen}}, \bibinfo {author} {\bibfnamefont
			{D.~F.}\ \bibnamefont {Price}}, \bibinfo {author} {\bibfnamefont
			{H.}~\bibnamefont {Ruhl}}, \bibinfo {author} {\bibfnamefont {P.~T.}\
			\bibnamefont {Springer}},\ and\ \bibinfo {author} {\bibfnamefont
			{R.}~\bibnamefont {Stephens}},\ }\bibfield  {title} {\bibinfo {title}
		{Isochoric heating of solid-density matter with an ultrafast proton beam},\
	}\href {https://doi.org/10.1103/PhysRevLett.91.125004} {\bibfield  {journal}
		{\bibinfo  {journal} {Phys. Rev. Lett.}\ }\textbf {\bibinfo {volume} {91}},\
		\bibinfo {pages} {125004} (\bibinfo {year} {2003})}\BibitemShut {NoStop}%
	\bibitem [{\citenamefont {Tabak}\ \emph {et~al.}(1994)\citenamefont {Tabak},
		\citenamefont {Hammer}, \citenamefont {Glinsky}, \citenamefont {Kruer},
		\citenamefont {Wilks}, \citenamefont {Woodworth}, \citenamefont {Campbell},
		\citenamefont {Perry},\ and\ \citenamefont {Mason}}]{10.1063/1.870664}%
	\BibitemOpen
	\bibfield  {author} {\bibinfo {author} {\bibfnamefont {M.}~\bibnamefont
			{Tabak}}, \bibinfo {author} {\bibfnamefont {J.}~\bibnamefont {Hammer}},
		\bibinfo {author} {\bibfnamefont {M.~E.}\ \bibnamefont {Glinsky}}, \bibinfo
		{author} {\bibfnamefont {W.~L.}\ \bibnamefont {Kruer}}, \bibinfo {author}
		{\bibfnamefont {S.~C.}\ \bibnamefont {Wilks}}, \bibinfo {author}
		{\bibfnamefont {J.}~\bibnamefont {Woodworth}}, \bibinfo {author}
		{\bibfnamefont {E.~M.}\ \bibnamefont {Campbell}}, \bibinfo {author}
		{\bibfnamefont {M.~D.}\ \bibnamefont {Perry}},\ and\ \bibinfo {author}
		{\bibfnamefont {R.~J.}\ \bibnamefont {Mason}},\ }\bibfield  {title} {\bibinfo
		{title} {Ignition and high gain with ultrapowerful lasers*},\ }\href
	{https://doi.org/10.1063/1.870664} {\bibfield  {journal} {\bibinfo  {journal}
			{Phys. Plasmas}\ }\textbf {\bibinfo {volume} {1}},\ \bibinfo {pages} {1626}
		(\bibinfo {year} {1994})}\BibitemShut {NoStop}%
	\bibitem [{\citenamefont {Naumova}\ \emph {et~al.}(2009)\citenamefont
		{Naumova}, \citenamefont {Schlegel}, \citenamefont {Tikhonchuk},
		\citenamefont {Labaune}, \citenamefont {Sokolov},\ and\ \citenamefont
		{Mourou}}]{PhysRevLett.102.025002}%
	\BibitemOpen
	\bibfield  {author} {\bibinfo {author} {\bibfnamefont {N.}~\bibnamefont
			{Naumova}}, \bibinfo {author} {\bibfnamefont {T.}~\bibnamefont {Schlegel}},
		\bibinfo {author} {\bibfnamefont {V.~T.}\ \bibnamefont {Tikhonchuk}},
		\bibinfo {author} {\bibfnamefont {C.}~\bibnamefont {Labaune}}, \bibinfo
		{author} {\bibfnamefont {I.~V.}\ \bibnamefont {Sokolov}},\ and\ \bibinfo
		{author} {\bibfnamefont {G.}~\bibnamefont {Mourou}},\ }\bibfield  {title}
	{\bibinfo {title} {Hole boring in a dt pellet and fast-ion ignition with
			ultraintense laser pulses},\ }\href
	{https://doi.org/10.1103/PhysRevLett.102.025002} {\bibfield  {journal}
		{\bibinfo  {journal} {Phys. Rev. Lett.}\ }\textbf {\bibinfo {volume} {102}},\
		\bibinfo {pages} {025002} (\bibinfo {year} {2009})}\BibitemShut {NoStop}%
	\bibitem [{\citenamefont {Krushelnick}\ \emph {et~al.}(2000)\citenamefont
		{Krushelnick}, \citenamefont {Clark}, \citenamefont {Allott}, \citenamefont
		{Beg}, \citenamefont {Danson}, \citenamefont {Machacek}, \citenamefont
		{Malka}, \citenamefont {Najmudin}, \citenamefont {Neely}, \citenamefont
		{Norreys}, \citenamefont {Salvati}, \citenamefont {Santala}, \citenamefont
		{Tatarakis}, \citenamefont {Watts}, \citenamefont {Zepf},\ and\ \citenamefont
		{Dangor}}]{893296}%
	\BibitemOpen
	\bibfield  {author} {\bibinfo {author} {\bibfnamefont {K.}~\bibnamefont
			{Krushelnick}}, \bibinfo {author} {\bibfnamefont {E.}~\bibnamefont {Clark}},
		\bibinfo {author} {\bibfnamefont {R.}~\bibnamefont {Allott}}, \bibinfo
		{author} {\bibfnamefont {F.}~\bibnamefont {Beg}}, \bibinfo {author}
		{\bibfnamefont {C.}~\bibnamefont {Danson}}, \bibinfo {author} {\bibfnamefont
			{A.}~\bibnamefont {Machacek}}, \bibinfo {author} {\bibfnamefont
			{V.}~\bibnamefont {Malka}}, \bibinfo {author} {\bibfnamefont
			{Z.}~\bibnamefont {Najmudin}}, \bibinfo {author} {\bibfnamefont
			{D.}~\bibnamefont {Neely}}, \bibinfo {author} {\bibfnamefont
			{P.}~\bibnamefont {Norreys}}, \bibinfo {author} {\bibfnamefont
			{M.}~\bibnamefont {Salvati}}, \bibinfo {author} {\bibfnamefont
			{M.}~\bibnamefont {Santala}}, \bibinfo {author} {\bibfnamefont
			{M.}~\bibnamefont {Tatarakis}}, \bibinfo {author} {\bibfnamefont
			{I.}~\bibnamefont {Watts}}, \bibinfo {author} {\bibfnamefont
			{M.}~\bibnamefont {Zepf}},\ and\ \bibinfo {author} {\bibfnamefont
			{A.}~\bibnamefont {Dangor}},\ }\bibfield  {title} {\bibinfo {title}
		{Ultrahigh-intensity laser-produced plasmas as a compact heavy ion injection
			source},\ }\href {https://doi.org/10.1109/27.893296} {\bibfield  {journal}
		{\bibinfo  {journal} {IEEE Trans. Plasma Sci.}\ }\textbf {\bibinfo {volume}
			{28}},\ \bibinfo {pages} {1110} (\bibinfo {year} {2000})}\BibitemShut
	{NoStop}%
	\bibitem [{\citenamefont {Gitomer}\ \emph {et~al.}(1986)\citenamefont
		{Gitomer}, \citenamefont {Jones}, \citenamefont {Begay}, \citenamefont
		{Ehler}, \citenamefont {Kephart},\ and\ \citenamefont
		{Kristal}}]{10.1063/1.865510}%
	\BibitemOpen
	\bibfield  {author} {\bibinfo {author} {\bibfnamefont {S.~J.}\ \bibnamefont
			{Gitomer}}, \bibinfo {author} {\bibfnamefont {R.~D.}\ \bibnamefont {Jones}},
		\bibinfo {author} {\bibfnamefont {F.}~\bibnamefont {Begay}}, \bibinfo
		{author} {\bibfnamefont {A.~W.}\ \bibnamefont {Ehler}}, \bibinfo {author}
		{\bibfnamefont {J.~F.}\ \bibnamefont {Kephart}},\ and\ \bibinfo {author}
		{\bibfnamefont {R.}~\bibnamefont {Kristal}},\ }\bibfield  {title} {\bibinfo
		{title} {Fast ions and hot electrons in the laser–plasma interaction},\
	}\href {https://doi.org/10.1063/1.865510} {\bibfield  {journal} {\bibinfo
			{journal} {Phys. Fluids}\ }\textbf {\bibinfo {volume} {29}},\ \bibinfo
		{pages} {2679} (\bibinfo {year} {1986})}\BibitemShut {NoStop}%
	\bibitem [{\citenamefont {Fews}\ \emph {et~al.}(1994)\citenamefont {Fews},
		\citenamefont {Norreys}, \citenamefont {Beg}, \citenamefont {Bell},
		\citenamefont {Dangor}, \citenamefont {Danson}, \citenamefont {Lee},\ and\
		\citenamefont {Rose}}]{PhysRevLett.73.1801}%
	\BibitemOpen
	\bibfield  {author} {\bibinfo {author} {\bibfnamefont {A.~P.}\ \bibnamefont
			{Fews}}, \bibinfo {author} {\bibfnamefont {P.~A.}\ \bibnamefont {Norreys}},
		\bibinfo {author} {\bibfnamefont {F.~N.}\ \bibnamefont {Beg}}, \bibinfo
		{author} {\bibfnamefont {A.~R.}\ \bibnamefont {Bell}}, \bibinfo {author}
		{\bibfnamefont {A.~E.}\ \bibnamefont {Dangor}}, \bibinfo {author}
		{\bibfnamefont {C.~N.}\ \bibnamefont {Danson}}, \bibinfo {author}
		{\bibfnamefont {P.}~\bibnamefont {Lee}},\ and\ \bibinfo {author}
		{\bibfnamefont {S.~J.}\ \bibnamefont {Rose}},\ }\bibfield  {title} {\bibinfo
		{title} {Plasma ion emission from high intensity picosecond laser pulse
			interactions with solid targets},\ }\href
	{https://doi.org/10.1103/PhysRevLett.73.1801} {\bibfield  {journal} {\bibinfo
			{journal} {Phys. Rev. Lett.}\ }\textbf {\bibinfo {volume} {73}},\ \bibinfo
		{pages} {1801} (\bibinfo {year} {1994})}\BibitemShut {NoStop}%
	\bibitem [{\citenamefont {Beg}\ \emph {et~al.}(1997)\citenamefont {Beg},
		\citenamefont {Bell}, \citenamefont {Dangor}, \citenamefont {Danson},
		\citenamefont {Fews}, \citenamefont {Glinsky}, \citenamefont {Hammel},
		\citenamefont {Lee}, \citenamefont {Norreys},\ and\ \citenamefont
		{Tatarakis}}]{10.1063/1.872103}%
	\BibitemOpen
	\bibfield  {author} {\bibinfo {author} {\bibfnamefont {F.~N.}\ \bibnamefont
			{Beg}}, \bibinfo {author} {\bibfnamefont {A.~R.}\ \bibnamefont {Bell}},
		\bibinfo {author} {\bibfnamefont {A.~E.}\ \bibnamefont {Dangor}}, \bibinfo
		{author} {\bibfnamefont {C.~N.}\ \bibnamefont {Danson}}, \bibinfo {author}
		{\bibfnamefont {A.~P.}\ \bibnamefont {Fews}}, \bibinfo {author}
		{\bibfnamefont {M.~E.}\ \bibnamefont {Glinsky}}, \bibinfo {author}
		{\bibfnamefont {B.~A.}\ \bibnamefont {Hammel}}, \bibinfo {author}
		{\bibfnamefont {P.}~\bibnamefont {Lee}}, \bibinfo {author} {\bibfnamefont
			{P.~A.}\ \bibnamefont {Norreys}},\ and\ \bibinfo {author} {\bibfnamefont
			{M.}~\bibnamefont {Tatarakis}},\ }\bibfield  {title} {\bibinfo {title} {A
			study of picosecond laser–solid interactions up to $10^{19}$ \text{W}
			cm$^{-2}$},\ }\href {https://doi.org/10.1063/1.872103} {\bibfield
		{journal} {\bibinfo  {journal} {Phys. Plasmas}\ }\textbf {\bibinfo {volume}
			{4}},\ \bibinfo {pages} {447} (\bibinfo {year} {1997})}\BibitemShut {NoStop}%
	\bibitem [{\citenamefont {Krushelnick}\ \emph {et~al.}(1999)\citenamefont
		{Krushelnick}, \citenamefont {Clark}, \citenamefont {Najmudin}, \citenamefont
		{Salvati}, \citenamefont {Santala}, \citenamefont {Tatarakis}, \citenamefont
		{Dangor}, \citenamefont {Malka}, \citenamefont {Neely}, \citenamefont
		{Allott},\ and\ \citenamefont {Danson}}]{PhysRevLett.83.737}%
	\BibitemOpen
	\bibfield  {author} {\bibinfo {author} {\bibfnamefont {K.}~\bibnamefont
			{Krushelnick}}, \bibinfo {author} {\bibfnamefont {E.~L.}\ \bibnamefont
			{Clark}}, \bibinfo {author} {\bibfnamefont {Z.}~\bibnamefont {Najmudin}},
		\bibinfo {author} {\bibfnamefont {M.}~\bibnamefont {Salvati}}, \bibinfo
		{author} {\bibfnamefont {M.~I.~K.}\ \bibnamefont {Santala}}, \bibinfo
		{author} {\bibfnamefont {M.}~\bibnamefont {Tatarakis}}, \bibinfo {author}
		{\bibfnamefont {A.~E.}\ \bibnamefont {Dangor}}, \bibinfo {author}
		{\bibfnamefont {V.}~\bibnamefont {Malka}}, \bibinfo {author} {\bibfnamefont
			{D.}~\bibnamefont {Neely}}, \bibinfo {author} {\bibfnamefont
			{R.}~\bibnamefont {Allott}},\ and\ \bibinfo {author} {\bibfnamefont
			{C.}~\bibnamefont {Danson}},\ }\bibfield  {title} {\bibinfo {title}
		{Multi-mev ion production from high-intensity laser interactions with
			underdense plasmas},\ }\href {https://doi.org/10.1103/PhysRevLett.83.737}
	{\bibfield  {journal} {\bibinfo  {journal} {Phys. Rev. Lett.}\ }\textbf
		{\bibinfo {volume} {83}},\ \bibinfo {pages} {737} (\bibinfo {year}
		{1999})}\BibitemShut {NoStop}%
	\bibitem [{\citenamefont {Sarkisov}\ \emph {et~al.}(1999)\citenamefont
		{Sarkisov}, \citenamefont {Bychenkov}, \citenamefont {Novikov}, \citenamefont
		{Tikhonchuk}, \citenamefont {Maksimchuk}, \citenamefont {Chen}, \citenamefont
		{Wagner}, \citenamefont {Mourou},\ and\ \citenamefont
		{Umstadter}}]{PhysRevE.59.7042}%
	\BibitemOpen
	\bibfield  {author} {\bibinfo {author} {\bibfnamefont {G.~S.}\ \bibnamefont
			{Sarkisov}}, \bibinfo {author} {\bibfnamefont {V.~Y.}\ \bibnamefont
			{Bychenkov}}, \bibinfo {author} {\bibfnamefont {V.~N.}\ \bibnamefont
			{Novikov}}, \bibinfo {author} {\bibfnamefont {V.~T.}\ \bibnamefont
			{Tikhonchuk}}, \bibinfo {author} {\bibfnamefont {A.}~\bibnamefont
			{Maksimchuk}}, \bibinfo {author} {\bibfnamefont {S.-Y.}\ \bibnamefont
			{Chen}}, \bibinfo {author} {\bibfnamefont {R.}~\bibnamefont {Wagner}},
		\bibinfo {author} {\bibfnamefont {G.}~\bibnamefont {Mourou}},\ and\ \bibinfo
		{author} {\bibfnamefont {D.}~\bibnamefont {Umstadter}},\ }\bibfield  {title}
	{\bibinfo {title} {Self-focusing, channel formation, and high-energy ion
			generation in interaction of an intense short laser pulse with a he jet},\
	}\href {https://doi.org/10.1103/PhysRevE.59.7042} {\bibfield  {journal}
		{\bibinfo  {journal} {Phys. Rev. E}\ }\textbf {\bibinfo {volume} {59}},\
		\bibinfo {pages} {7042} (\bibinfo {year} {1999})}\BibitemShut {NoStop}%
	\bibitem [{\citenamefont {Ditmire}\ \emph {et~al.}(1997)\citenamefont
		{Ditmire}, \citenamefont {Tisch}, \citenamefont {Springate}, \citenamefont
		{Mason}, \citenamefont {Hay}, \citenamefont {Smith}, \citenamefont
		{Marangos},\ and\ \citenamefont {Hutchinson}}]{Ditmire1997}%
	\BibitemOpen
	\bibfield  {author} {\bibinfo {author} {\bibfnamefont {T.}~\bibnamefont
			{Ditmire}}, \bibinfo {author} {\bibfnamefont {J.~W.~G.}\ \bibnamefont
			{Tisch}}, \bibinfo {author} {\bibfnamefont {E.}~\bibnamefont {Springate}},
		\bibinfo {author} {\bibfnamefont {M.~B.}\ \bibnamefont {Mason}}, \bibinfo
		{author} {\bibfnamefont {N.}~\bibnamefont {Hay}}, \bibinfo {author}
		{\bibfnamefont {R.~A.}\ \bibnamefont {Smith}}, \bibinfo {author}
		{\bibfnamefont {J.}~\bibnamefont {Marangos}},\ and\ \bibinfo {author}
		{\bibfnamefont {M.~H.~R.}\ \bibnamefont {Hutchinson}},\ }\bibfield  {title}
	{\bibinfo {title} {High-energy ions produced in explosions of superheated
			atomic clusters},\ }\href {https://doi.org/10.1038/386054a0} {\bibfield
		{journal} {\bibinfo  {journal} {Nature}\ }\textbf {\bibinfo {volume} {386}},\
		\bibinfo {pages} {54} (\bibinfo {year} {1997})}\BibitemShut {NoStop}%
	\bibitem [{\citenamefont {Clark}\ \emph {et~al.}(2000)\citenamefont {Clark},
		\citenamefont {Krushelnick}, \citenamefont {Davies}, \citenamefont {Zepf},
		\citenamefont {Tatarakis}, \citenamefont {Beg}, \citenamefont {Machacek},
		\citenamefont {Norreys}, \citenamefont {Santala}, \citenamefont {Watts},\
		and\ \citenamefont {Dangor}}]{PhysRevLett.84.670}%
	\BibitemOpen
	\bibfield  {author} {\bibinfo {author} {\bibfnamefont {E.~L.}\ \bibnamefont
			{Clark}}, \bibinfo {author} {\bibfnamefont {K.}~\bibnamefont {Krushelnick}},
		\bibinfo {author} {\bibfnamefont {J.~R.}\ \bibnamefont {Davies}}, \bibinfo
		{author} {\bibfnamefont {M.}~\bibnamefont {Zepf}}, \bibinfo {author}
		{\bibfnamefont {M.}~\bibnamefont {Tatarakis}}, \bibinfo {author}
		{\bibfnamefont {F.~N.}\ \bibnamefont {Beg}}, \bibinfo {author} {\bibfnamefont
			{A.}~\bibnamefont {Machacek}}, \bibinfo {author} {\bibfnamefont {P.~A.}\
			\bibnamefont {Norreys}}, \bibinfo {author} {\bibfnamefont {M.~I.~K.}\
			\bibnamefont {Santala}}, \bibinfo {author} {\bibfnamefont {I.}~\bibnamefont
			{Watts}},\ and\ \bibinfo {author} {\bibfnamefont {A.~E.}\ \bibnamefont
			{Dangor}},\ }\bibfield  {title} {\bibinfo {title} {Measurements of energetic
			proton transport through magnetized plasma from intense laser interactions
			with solids},\ }\href {https://doi.org/10.1103/PhysRevLett.84.670} {\bibfield
		{journal} {\bibinfo  {journal} {Phys. Rev. Lett.}\ }\textbf {\bibinfo
			{volume} {84}},\ \bibinfo {pages} {670} (\bibinfo {year} {2000})}\BibitemShut
	{NoStop}%
	\bibitem [{\citenamefont {Maksimchuk}\ \emph {et~al.}(2000)\citenamefont
		{Maksimchuk}, \citenamefont {Gu}, \citenamefont {Flippo}, \citenamefont
		{Umstadter},\ and\ \citenamefont {Bychenkov}}]{PhysRevLett.84.4108}%
	\BibitemOpen
	\bibfield  {author} {\bibinfo {author} {\bibfnamefont {A.}~\bibnamefont
			{Maksimchuk}}, \bibinfo {author} {\bibfnamefont {S.}~\bibnamefont {Gu}},
		\bibinfo {author} {\bibfnamefont {K.}~\bibnamefont {Flippo}}, \bibinfo
		{author} {\bibfnamefont {D.}~\bibnamefont {Umstadter}},\ and\ \bibinfo
		{author} {\bibfnamefont {V.~Y.}\ \bibnamefont {Bychenkov}},\ }\bibfield
	{title} {\bibinfo {title} {Forward ion acceleration in thin films driven by a
			high-intensity laser},\ }\href {https://doi.org/10.1103/PhysRevLett.84.4108}
	{\bibfield  {journal} {\bibinfo  {journal} {Phys. Rev. Lett.}\ }\textbf
		{\bibinfo {volume} {84}},\ \bibinfo {pages} {4108} (\bibinfo {year}
		{2000})}\BibitemShut {NoStop}%
	\bibitem [{\citenamefont {Snavely}\ \emph {et~al.}(2000)\citenamefont
		{Snavely}, \citenamefont {Key}, \citenamefont {Hatchett}, \citenamefont
		{Cowan}, \citenamefont {Roth}, \citenamefont {Phillips}, \citenamefont
		{Stoyer}, \citenamefont {Henry}, \citenamefont {Sangster}, \citenamefont
		{Singh}, \citenamefont {Wilks}, \citenamefont {MacKinnon}, \citenamefont
		{Offenberger}, \citenamefont {Pennington}, \citenamefont {Yasuike},
		\citenamefont {Langdon}, \citenamefont {Lasinski}, \citenamefont {Johnson},
		\citenamefont {Perry},\ and\ \citenamefont {Campbell}}]{PhysRevLett.85.2945}%
	\BibitemOpen
	\bibfield  {author} {\bibinfo {author} {\bibfnamefont {R.~A.}\ \bibnamefont
			{Snavely}}, \bibinfo {author} {\bibfnamefont {M.~H.}\ \bibnamefont {Key}},
		\bibinfo {author} {\bibfnamefont {S.~P.}\ \bibnamefont {Hatchett}}, \bibinfo
		{author} {\bibfnamefont {T.~E.}\ \bibnamefont {Cowan}}, \bibinfo {author}
		{\bibfnamefont {M.}~\bibnamefont {Roth}}, \bibinfo {author} {\bibfnamefont
			{T.~W.}\ \bibnamefont {Phillips}}, \bibinfo {author} {\bibfnamefont {M.~A.}\
			\bibnamefont {Stoyer}}, \bibinfo {author} {\bibfnamefont {E.~A.}\
			\bibnamefont {Henry}}, \bibinfo {author} {\bibfnamefont {T.~C.}\ \bibnamefont
			{Sangster}}, \bibinfo {author} {\bibfnamefont {M.~S.}\ \bibnamefont {Singh}},
		\bibinfo {author} {\bibfnamefont {S.~C.}\ \bibnamefont {Wilks}}, \bibinfo
		{author} {\bibfnamefont {A.}~\bibnamefont {MacKinnon}}, \bibinfo {author}
		{\bibfnamefont {A.}~\bibnamefont {Offenberger}}, \bibinfo {author}
		{\bibfnamefont {D.~M.}\ \bibnamefont {Pennington}}, \bibinfo {author}
		{\bibfnamefont {K.}~\bibnamefont {Yasuike}}, \bibinfo {author} {\bibfnamefont
			{A.~B.}\ \bibnamefont {Langdon}}, \bibinfo {author} {\bibfnamefont {B.~F.}\
			\bibnamefont {Lasinski}}, \bibinfo {author} {\bibfnamefont {J.}~\bibnamefont
			{Johnson}}, \bibinfo {author} {\bibfnamefont {M.~D.}\ \bibnamefont {Perry}},\
		and\ \bibinfo {author} {\bibfnamefont {E.~M.}\ \bibnamefont {Campbell}},\
	}\bibfield  {title} {\bibinfo {title} {Intense high-energy proton beams from
			petawatt-laser irradiation of solids},\ }\href
	{https://doi.org/10.1103/PhysRevLett.85.2945} {\bibfield  {journal} {\bibinfo
			{journal} {Phys. Rev. Lett.}\ }\textbf {\bibinfo {volume} {85}},\ \bibinfo
		{pages} {2945} (\bibinfo {year} {2000})}\BibitemShut {NoStop}%
	\bibitem [{\citenamefont {Wilks}\ \emph {et~al.}(2001)\citenamefont {Wilks},
		\citenamefont {Langdon}, \citenamefont {Cowan}, \citenamefont {Roth},
		\citenamefont {Singh}, \citenamefont {Hatchett}, \citenamefont {Key},
		\citenamefont {Pennington}, \citenamefont {MacKinnon},\ and\ \citenamefont
		{Snavely}}]{10.1063/1.1333697}%
	\BibitemOpen
	\bibfield  {author} {\bibinfo {author} {\bibfnamefont {S.~C.}\ \bibnamefont
			{Wilks}}, \bibinfo {author} {\bibfnamefont {A.~B.}\ \bibnamefont {Langdon}},
		\bibinfo {author} {\bibfnamefont {T.~E.}\ \bibnamefont {Cowan}}, \bibinfo
		{author} {\bibfnamefont {M.}~\bibnamefont {Roth}}, \bibinfo {author}
		{\bibfnamefont {M.}~\bibnamefont {Singh}}, \bibinfo {author} {\bibfnamefont
			{S.}~\bibnamefont {Hatchett}}, \bibinfo {author} {\bibfnamefont {M.~H.}\
			\bibnamefont {Key}}, \bibinfo {author} {\bibfnamefont {D.}~\bibnamefont
			{Pennington}}, \bibinfo {author} {\bibfnamefont {A.}~\bibnamefont
			{MacKinnon}},\ and\ \bibinfo {author} {\bibfnamefont {R.~A.}\ \bibnamefont
			{Snavely}},\ }\bibfield  {title} {\bibinfo {title} {Energetic proton
			generation in ultra-intense laser–solid interactions},\ }\href
	{https://doi.org/10.1063/1.1333697} {\bibfield  {journal} {\bibinfo
			{journal} {Phys. Plasmas}\ }\textbf {\bibinfo {volume} {8}},\ \bibinfo
		{pages} {542} (\bibinfo {year} {2001})}\BibitemShut {NoStop}%
	\bibitem [{\citenamefont {Borghesi}\ \emph {et~al.}(2004)\citenamefont
		{Borghesi}, \citenamefont {Mackinnon}, \citenamefont {Campbell},
		\citenamefont {Hicks}, \citenamefont {Kar}, \citenamefont {Patel},
		\citenamefont {Price}, \citenamefont {Romagnani}, \citenamefont {Schiavi},\
		and\ \citenamefont {Willi}}]{PhysRevLett.92.055003}%
	\BibitemOpen
	\bibfield  {author} {\bibinfo {author} {\bibfnamefont {M.}~\bibnamefont
			{Borghesi}}, \bibinfo {author} {\bibfnamefont {A.~J.}\ \bibnamefont
			{Mackinnon}}, \bibinfo {author} {\bibfnamefont {D.~H.}\ \bibnamefont
			{Campbell}}, \bibinfo {author} {\bibfnamefont {D.~G.}\ \bibnamefont {Hicks}},
		\bibinfo {author} {\bibfnamefont {S.}~\bibnamefont {Kar}}, \bibinfo {author}
		{\bibfnamefont {P.~K.}\ \bibnamefont {Patel}}, \bibinfo {author}
		{\bibfnamefont {D.}~\bibnamefont {Price}}, \bibinfo {author} {\bibfnamefont
			{L.}~\bibnamefont {Romagnani}}, \bibinfo {author} {\bibfnamefont
			{A.}~\bibnamefont {Schiavi}},\ and\ \bibinfo {author} {\bibfnamefont
			{O.}~\bibnamefont {Willi}},\ }\bibfield  {title} {\bibinfo {title} {Multi-mev
			proton source investigations in ultraintense laser-foil interactions},\
	}\href {https://doi.org/10.1103/PhysRevLett.92.055003} {\bibfield  {journal}
		{\bibinfo  {journal} {Phys. Rev. Lett.}\ }\textbf {\bibinfo {volume} {92}},\
		\bibinfo {pages} {055003} (\bibinfo {year} {2004})}\BibitemShut {NoStop}%
	\bibitem [{\citenamefont {Zepf}\ \emph {et~al.}(2003)\citenamefont {Zepf},
		\citenamefont {Clark}, \citenamefont {Beg}, \citenamefont {Clarke},
		\citenamefont {Dangor}, \citenamefont {Gopal}, \citenamefont {Krushelnick},
		\citenamefont {Norreys}, \citenamefont {Tatarakis}, \citenamefont {Wagner},\
		and\ \citenamefont {Wei}}]{PhysRevLett.90.064801}%
	\BibitemOpen
	\bibfield  {author} {\bibinfo {author} {\bibfnamefont {M.}~\bibnamefont
			{Zepf}}, \bibinfo {author} {\bibfnamefont {E.~L.}\ \bibnamefont {Clark}},
		\bibinfo {author} {\bibfnamefont {F.~N.}\ \bibnamefont {Beg}}, \bibinfo
		{author} {\bibfnamefont {R.~J.}\ \bibnamefont {Clarke}}, \bibinfo {author}
		{\bibfnamefont {A.~E.}\ \bibnamefont {Dangor}}, \bibinfo {author}
		{\bibfnamefont {A.}~\bibnamefont {Gopal}}, \bibinfo {author} {\bibfnamefont
			{K.}~\bibnamefont {Krushelnick}}, \bibinfo {author} {\bibfnamefont {P.~A.}\
			\bibnamefont {Norreys}}, \bibinfo {author} {\bibfnamefont {M.}~\bibnamefont
			{Tatarakis}}, \bibinfo {author} {\bibfnamefont {U.}~\bibnamefont {Wagner}},\
		and\ \bibinfo {author} {\bibfnamefont {M.~S.}\ \bibnamefont {Wei}},\
	}\bibfield  {title} {\bibinfo {title} {Proton acceleration from
			high-intensity laser interactions with thin foil targets},\ }\href
	{https://doi.org/10.1103/PhysRevLett.90.064801} {\bibfield  {journal}
		{\bibinfo  {journal} {Phys. Rev. Lett.}\ }\textbf {\bibinfo {volume} {90}},\
		\bibinfo {pages} {064801} (\bibinfo {year} {2003})}\BibitemShut {NoStop}%
	\bibitem [{\citenamefont {Cowan}\ \emph {et~al.}(2004)\citenamefont {Cowan},
		\citenamefont {Fuchs}, \citenamefont {Ruhl}, \citenamefont {Kemp},
		\citenamefont {Audebert}, \citenamefont {Roth}, \citenamefont {Stephens},
		\citenamefont {Barton}, \citenamefont {Blazevic}, \citenamefont {Brambrink},
		\citenamefont {Cobble}, \citenamefont {Fern\'andez}, \citenamefont
		{Gauthier}, \citenamefont {Geissel}, \citenamefont {Hegelich}, \citenamefont
		{Kaae}, \citenamefont {Karsch}, \citenamefont {Le~Sage}, \citenamefont
		{Letzring}, \citenamefont {Manclossi}, \citenamefont {Meyroneinc},
		\citenamefont {Newkirk}, \citenamefont {P\'epin},\ and\ \citenamefont
		{Renard-LeGalloudec}}]{PhysRevLett.92.204801}%
	\BibitemOpen
	\bibfield  {author} {\bibinfo {author} {\bibfnamefont {T.~E.}\ \bibnamefont
			{Cowan}}, \bibinfo {author} {\bibfnamefont {J.}~\bibnamefont {Fuchs}},
		\bibinfo {author} {\bibfnamefont {H.}~\bibnamefont {Ruhl}}, \bibinfo {author}
		{\bibfnamefont {A.}~\bibnamefont {Kemp}}, \bibinfo {author} {\bibfnamefont
			{P.}~\bibnamefont {Audebert}}, \bibinfo {author} {\bibfnamefont
			{M.}~\bibnamefont {Roth}}, \bibinfo {author} {\bibfnamefont {R.}~\bibnamefont
			{Stephens}}, \bibinfo {author} {\bibfnamefont {I.}~\bibnamefont {Barton}},
		\bibinfo {author} {\bibfnamefont {A.}~\bibnamefont {Blazevic}}, \bibinfo
		{author} {\bibfnamefont {E.}~\bibnamefont {Brambrink}}, \bibinfo {author}
		{\bibfnamefont {J.}~\bibnamefont {Cobble}}, \bibinfo {author} {\bibfnamefont
			{J.}~\bibnamefont {Fern\'andez}}, \bibinfo {author} {\bibfnamefont {J.-C.}\
			\bibnamefont {Gauthier}}, \bibinfo {author} {\bibfnamefont {M.}~\bibnamefont
			{Geissel}}, \bibinfo {author} {\bibfnamefont {M.}~\bibnamefont {Hegelich}},
		\bibinfo {author} {\bibfnamefont {J.}~\bibnamefont {Kaae}}, \bibinfo {author}
		{\bibfnamefont {S.}~\bibnamefont {Karsch}}, \bibinfo {author} {\bibfnamefont
			{G.~P.}\ \bibnamefont {Le~Sage}}, \bibinfo {author} {\bibfnamefont
			{S.}~\bibnamefont {Letzring}}, \bibinfo {author} {\bibfnamefont
			{M.}~\bibnamefont {Manclossi}}, \bibinfo {author} {\bibfnamefont
			{S.}~\bibnamefont {Meyroneinc}}, \bibinfo {author} {\bibfnamefont
			{A.}~\bibnamefont {Newkirk}}, \bibinfo {author} {\bibfnamefont
			{H.}~\bibnamefont {P\'epin}},\ and\ \bibinfo {author} {\bibfnamefont
			{N.}~\bibnamefont {Renard-LeGalloudec}},\ }\bibfield  {title} {\bibinfo
		{title} {Ultralow emittance, multi-mev proton beams from a laser
			virtual-cathode plasma accelerator},\ }\href
	{https://doi.org/10.1103/PhysRevLett.92.204801} {\bibfield  {journal}
		{\bibinfo  {journal} {Phys. Rev. Lett.}\ }\textbf {\bibinfo {volume} {92}},\
		\bibinfo {pages} {204801} (\bibinfo {year} {2004})}\BibitemShut {NoStop}%
	\bibitem [{\citenamefont {Nürnberg}\ \emph {et~al.}(2009)\citenamefont
		{Nürnberg}, \citenamefont {Schollmeier}, \citenamefont {Brambrink},
		\citenamefont {Blažević}, \citenamefont {Carroll}, \citenamefont {Flippo},
		\citenamefont {Gautier}, \citenamefont {Geißel}, \citenamefont {Harres},
		\citenamefont {Hegelich}, \citenamefont {Lundh}, \citenamefont {Markey},
		\citenamefont {McKenna}, \citenamefont {Neely}, \citenamefont {Schreiber},\
		and\ \citenamefont {Roth}}]{10.1063/1.3086424}%
	\BibitemOpen
	\bibfield  {author} {\bibinfo {author} {\bibfnamefont {F.}~\bibnamefont
			{Nürnberg}}, \bibinfo {author} {\bibfnamefont {M.}~\bibnamefont
			{Schollmeier}}, \bibinfo {author} {\bibfnamefont {E.}~\bibnamefont
			{Brambrink}}, \bibinfo {author} {\bibfnamefont {A.}~\bibnamefont
			{Blažević}}, \bibinfo {author} {\bibfnamefont {D.~C.}\ \bibnamefont
			{Carroll}}, \bibinfo {author} {\bibfnamefont {K.}~\bibnamefont {Flippo}},
		\bibinfo {author} {\bibfnamefont {D.~C.}\ \bibnamefont {Gautier}}, \bibinfo
		{author} {\bibfnamefont {M.}~\bibnamefont {Geißel}}, \bibinfo {author}
		{\bibfnamefont {K.}~\bibnamefont {Harres}}, \bibinfo {author} {\bibfnamefont
			{B.~M.}\ \bibnamefont {Hegelich}}, \bibinfo {author} {\bibfnamefont
			{O.}~\bibnamefont {Lundh}}, \bibinfo {author} {\bibfnamefont
			{K.}~\bibnamefont {Markey}}, \bibinfo {author} {\bibfnamefont
			{P.}~\bibnamefont {McKenna}}, \bibinfo {author} {\bibfnamefont
			{D.}~\bibnamefont {Neely}}, \bibinfo {author} {\bibfnamefont
			{J.}~\bibnamefont {Schreiber}},\ and\ \bibinfo {author} {\bibfnamefont
			{M.}~\bibnamefont {Roth}},\ }\bibfield  {title} {\bibinfo {title}
		{Radiochromic film imaging spectroscopy of laser-accelerated proton beams},\
	}\href {https://doi.org/10.1063/1.3086424} {\bibfield  {journal} {\bibinfo
			{journal} {Rev. Sci. Instrum.}\ }\textbf {\bibinfo {volume} {80}},\ \bibinfo
		{pages} {033301} (\bibinfo {year} {2009})}\BibitemShut {NoStop}%
	\bibitem [{\citenamefont {Ruhl}\ \emph {et~al.}(2001)\citenamefont {Ruhl},
		\citenamefont {Bulanov}, \citenamefont {Cowan}, \citenamefont {Liseĭkina},
		\citenamefont {Nickles}, \citenamefont {Pegoraro}, \citenamefont {Roth},\
		and\ \citenamefont {Sandner}}]{Ruhl2001}%
	\BibitemOpen
	\bibfield  {author} {\bibinfo {author} {\bibfnamefont {H.}~\bibnamefont
			{Ruhl}}, \bibinfo {author} {\bibfnamefont {S.~V.}\ \bibnamefont {Bulanov}},
		\bibinfo {author} {\bibfnamefont {T.~E.}\ \bibnamefont {Cowan}}, \bibinfo
		{author} {\bibfnamefont {T.~V.}\ \bibnamefont {Liseĭkina}}, \bibinfo
		{author} {\bibfnamefont {P.}~\bibnamefont {Nickles}}, \bibinfo {author}
		{\bibfnamefont {F.}~\bibnamefont {Pegoraro}}, \bibinfo {author}
		{\bibfnamefont {M.}~\bibnamefont {Roth}},\ and\ \bibinfo {author}
		{\bibfnamefont {W.}~\bibnamefont {Sandner}},\ }\bibfield  {title} {\bibinfo
		{title} {Computer simulation of the three-dimensional regime of proton
			acceleration in the interaction of laser radiation with a thin spherical
			target},\ }\href {https://doi.org/10.1134/1.1371596} {\bibfield  {journal}
		{\bibinfo  {journal} {Plasma Phys. Rep.}\ }\textbf {\bibinfo {volume} {27}},\
		\bibinfo {pages} {363} (\bibinfo {year} {2001})}\BibitemShut {NoStop}%
	\bibitem [{\citenamefont {Snavely}\ \emph {et~al.}(2007)\citenamefont
		{Snavely}, \citenamefont {Zhang}, \citenamefont {Akli}, \citenamefont {Chen},
		\citenamefont {Freeman}, \citenamefont {Gu}, \citenamefont {Hatchett},
		\citenamefont {Hey}, \citenamefont {Hill}, \citenamefont {Key}, \citenamefont
		{Izawa}, \citenamefont {King}, \citenamefont {Kitagawa}, \citenamefont
		{Kodama}, \citenamefont {Langdon}, \citenamefont {Lasinski}, \citenamefont
		{Lei}, \citenamefont {MacKinnon}, \citenamefont {Patel}, \citenamefont
		{Stephens}, \citenamefont {Tampo}, \citenamefont {Tanaka}, \citenamefont
		{Town}, \citenamefont {Toyama}, \citenamefont {Tsutsumi}, \citenamefont
		{Wilks}, \citenamefont {Yabuuchi},\ and\ \citenamefont
		{Zheng}}]{10.1063/1.2774001}%
	\BibitemOpen
	\bibfield  {author} {\bibinfo {author} {\bibfnamefont {R.~A.}\ \bibnamefont
			{Snavely}}, \bibinfo {author} {\bibfnamefont {B.}~\bibnamefont {Zhang}},
		\bibinfo {author} {\bibfnamefont {K.}~\bibnamefont {Akli}}, \bibinfo {author}
		{\bibfnamefont {Z.}~\bibnamefont {Chen}}, \bibinfo {author} {\bibfnamefont
			{R.~R.}\ \bibnamefont {Freeman}}, \bibinfo {author} {\bibfnamefont
			{P.}~\bibnamefont {Gu}}, \bibinfo {author} {\bibfnamefont {S.~P.}\
			\bibnamefont {Hatchett}}, \bibinfo {author} {\bibfnamefont {D.}~\bibnamefont
			{Hey}}, \bibinfo {author} {\bibfnamefont {J.}~\bibnamefont {Hill}}, \bibinfo
		{author} {\bibfnamefont {M.~H.}\ \bibnamefont {Key}}, \bibinfo {author}
		{\bibfnamefont {Y.}~\bibnamefont {Izawa}}, \bibinfo {author} {\bibfnamefont
			{J.}~\bibnamefont {King}}, \bibinfo {author} {\bibfnamefont {Y.}~\bibnamefont
			{Kitagawa}}, \bibinfo {author} {\bibfnamefont {R.}~\bibnamefont {Kodama}},
		\bibinfo {author} {\bibfnamefont {A.~B.}\ \bibnamefont {Langdon}}, \bibinfo
		{author} {\bibfnamefont {B.~F.}\ \bibnamefont {Lasinski}}, \bibinfo {author}
		{\bibfnamefont {A.}~\bibnamefont {Lei}}, \bibinfo {author} {\bibfnamefont
			{A.~J.}\ \bibnamefont {MacKinnon}}, \bibinfo {author} {\bibfnamefont
			{P.}~\bibnamefont {Patel}}, \bibinfo {author} {\bibfnamefont
			{R.}~\bibnamefont {Stephens}}, \bibinfo {author} {\bibfnamefont
			{M.}~\bibnamefont {Tampo}}, \bibinfo {author} {\bibfnamefont {K.~A.}\
			\bibnamefont {Tanaka}}, \bibinfo {author} {\bibfnamefont {R.}~\bibnamefont
			{Town}}, \bibinfo {author} {\bibfnamefont {Y.}~\bibnamefont {Toyama}},
		\bibinfo {author} {\bibfnamefont {T.}~\bibnamefont {Tsutsumi}}, \bibinfo
		{author} {\bibfnamefont {S.~C.}\ \bibnamefont {Wilks}}, \bibinfo {author}
		{\bibfnamefont {T.}~\bibnamefont {Yabuuchi}},\ and\ \bibinfo {author}
		{\bibfnamefont {J.}~\bibnamefont {Zheng}},\ }\bibfield  {title} {\bibinfo
		{title} {Laser generated proton beam focusing and high temperature isochoric
			heating of solid matter},\ }\href {https://doi.org/10.1063/1.2774001}
	{\bibfield  {journal} {\bibinfo  {journal} {Phys. Plasmas}\ }\textbf
		{\bibinfo {volume} {14}},\ \bibinfo {pages} {092703} (\bibinfo {year}
		{2007})}\BibitemShut {NoStop}%
	\bibitem [{\citenamefont {Kar}\ \emph {et~al.}(2011)\citenamefont {Kar},
		\citenamefont {Markey}, \citenamefont {Borghesi}, \citenamefont {Carroll},
		\citenamefont {McKenna}, \citenamefont {Neely}, \citenamefont {Quinn},\ and\
		\citenamefont {Zepf}}]{PhysRevLett.106.225003}%
	\BibitemOpen
	\bibfield  {author} {\bibinfo {author} {\bibfnamefont {S.}~\bibnamefont
			{Kar}}, \bibinfo {author} {\bibfnamefont {K.}~\bibnamefont {Markey}},
		\bibinfo {author} {\bibfnamefont {M.}~\bibnamefont {Borghesi}}, \bibinfo
		{author} {\bibfnamefont {D.~C.}\ \bibnamefont {Carroll}}, \bibinfo {author}
		{\bibfnamefont {P.}~\bibnamefont {McKenna}}, \bibinfo {author} {\bibfnamefont
			{D.}~\bibnamefont {Neely}}, \bibinfo {author} {\bibfnamefont {M.~N.}\
			\bibnamefont {Quinn}},\ and\ \bibinfo {author} {\bibfnamefont
			{M.}~\bibnamefont {Zepf}},\ }\bibfield  {title} {\bibinfo {title} {Ballistic
			focusing of polyenergetic protons driven by petawatt laser pulses},\ }\href
	{https://doi.org/10.1103/PhysRevLett.106.225003} {\bibfield  {journal}
		{\bibinfo  {journal} {Phys. Rev. Lett.}\ }\textbf {\bibinfo {volume} {106}},\
		\bibinfo {pages} {225003} (\bibinfo {year} {2011})}\BibitemShut {NoStop}%
	\bibitem [{\citenamefont {Bartal}\ \emph {et~al.}(2012)\citenamefont {Bartal},
		\citenamefont {Foord}, \citenamefont {Bellei}, \citenamefont {Key},
		\citenamefont {Flippo}, \citenamefont {Gaillard}, \citenamefont {Offermann},
		\citenamefont {Patel}, \citenamefont {Jarrott}, \citenamefont {Higginson},
		\citenamefont {Roth}, \citenamefont {Otten}, \citenamefont {Kraus},
		\citenamefont {Stephens}, \citenamefont {McLean}, \citenamefont {Giraldez},
		\citenamefont {Wei}, \citenamefont {Gautier},\ and\ \citenamefont
		{Beg}}]{Bartal2012}%
	\BibitemOpen
	\bibfield  {author} {\bibinfo {author} {\bibfnamefont {T.}~\bibnamefont
			{Bartal}}, \bibinfo {author} {\bibfnamefont {M.~E.}\ \bibnamefont {Foord}},
		\bibinfo {author} {\bibfnamefont {C.}~\bibnamefont {Bellei}}, \bibinfo
		{author} {\bibfnamefont {M.~H.}\ \bibnamefont {Key}}, \bibinfo {author}
		{\bibfnamefont {K.~A.}\ \bibnamefont {Flippo}}, \bibinfo {author}
		{\bibfnamefont {S.~A.}\ \bibnamefont {Gaillard}}, \bibinfo {author}
		{\bibfnamefont {D.~T.}\ \bibnamefont {Offermann}}, \bibinfo {author}
		{\bibfnamefont {P.~K.}\ \bibnamefont {Patel}}, \bibinfo {author}
		{\bibfnamefont {L.~C.}\ \bibnamefont {Jarrott}}, \bibinfo {author}
		{\bibfnamefont {D.~P.}\ \bibnamefont {Higginson}}, \bibinfo {author}
		{\bibfnamefont {M.}~\bibnamefont {Roth}}, \bibinfo {author} {\bibfnamefont
			{A.}~\bibnamefont {Otten}}, \bibinfo {author} {\bibfnamefont
			{D.}~\bibnamefont {Kraus}}, \bibinfo {author} {\bibfnamefont {R.~B.}\
			\bibnamefont {Stephens}}, \bibinfo {author} {\bibfnamefont {H.~S.}\
			\bibnamefont {McLean}}, \bibinfo {author} {\bibfnamefont {E.~M.}\
			\bibnamefont {Giraldez}}, \bibinfo {author} {\bibfnamefont {M.~S.}\
			\bibnamefont {Wei}}, \bibinfo {author} {\bibfnamefont {D.~C.}\ \bibnamefont
			{Gautier}},\ and\ \bibinfo {author} {\bibfnamefont {F.~N.}\ \bibnamefont
			{Beg}},\ }\bibfield  {title} {\bibinfo {title} {Focusing of short-pulse
			high-intensity laser-accelerated proton beams},\ }\href
	{https://doi.org/10.1038/nphys2153} {\bibfield  {journal} {\bibinfo
			{journal} {Nat. Phys.}\ }\textbf {\bibinfo {volume} {8}},\ \bibinfo {pages}
		{139} (\bibinfo {year} {2012})}\BibitemShut {NoStop}%
	\bibitem [{\citenamefont {Qiao}\ \emph {et~al.}(2013)\citenamefont {Qiao},
		\citenamefont {Foord}, \citenamefont {Wei}, \citenamefont {Stephens},
		\citenamefont {Key}, \citenamefont {McLean}, \citenamefont {Patel},\ and\
		\citenamefont {Beg}}]{PhysRevE.87.013108}%
	\BibitemOpen
	\bibfield  {author} {\bibinfo {author} {\bibfnamefont {B.}~\bibnamefont
			{Qiao}}, \bibinfo {author} {\bibfnamefont {M.~E.}\ \bibnamefont {Foord}},
		\bibinfo {author} {\bibfnamefont {M.~S.}\ \bibnamefont {Wei}}, \bibinfo
		{author} {\bibfnamefont {R.~B.}\ \bibnamefont {Stephens}}, \bibinfo {author}
		{\bibfnamefont {M.~H.}\ \bibnamefont {Key}}, \bibinfo {author} {\bibfnamefont
			{H.}~\bibnamefont {McLean}}, \bibinfo {author} {\bibfnamefont {P.~K.}\
			\bibnamefont {Patel}},\ and\ \bibinfo {author} {\bibfnamefont {F.~N.}\
			\bibnamefont {Beg}},\ }\bibfield  {title} {\bibinfo {title} {Dynamics of
			high-energy proton beam acceleration and focusing from hemisphere-cone
			targets by high-intensity lasers},\ }\href
	{https://doi.org/10.1103/PhysRevE.87.013108} {\bibfield  {journal} {\bibinfo
			{journal} {Phys. Rev. E}\ }\textbf {\bibinfo {volume} {87}},\ \bibinfo
		{pages} {013108} (\bibinfo {year} {2013})}\BibitemShut {NoStop}%
	\bibitem [{\citenamefont {McGuffey}\ \emph {et~al.}(2020)\citenamefont
		{McGuffey}, \citenamefont {Kim}, \citenamefont {Wei}, \citenamefont {Nilson},
		\citenamefont {Chen}, \citenamefont {Fuchs}, \citenamefont {Fitzsimmons},
		\citenamefont {Foord}, \citenamefont {Mariscal}, \citenamefont {McLean},
		\citenamefont {Patel}, \citenamefont {Stephens},\ and\ \citenamefont
		{Beg}}]{McGuffey2020}%
	\BibitemOpen
	\bibfield  {author} {\bibinfo {author} {\bibfnamefont {C.}~\bibnamefont
			{McGuffey}}, \bibinfo {author} {\bibfnamefont {J.}~\bibnamefont {Kim}},
		\bibinfo {author} {\bibfnamefont {M.~S.}\ \bibnamefont {Wei}}, \bibinfo
		{author} {\bibfnamefont {P.~M.}\ \bibnamefont {Nilson}}, \bibinfo {author}
		{\bibfnamefont {S.~N.}\ \bibnamefont {Chen}}, \bibinfo {author}
		{\bibfnamefont {J.}~\bibnamefont {Fuchs}}, \bibinfo {author} {\bibfnamefont
			{P.}~\bibnamefont {Fitzsimmons}}, \bibinfo {author} {\bibfnamefont {M.~E.}\
			\bibnamefont {Foord}}, \bibinfo {author} {\bibfnamefont {D.}~\bibnamefont
			{Mariscal}}, \bibinfo {author} {\bibfnamefont {H.~S.}\ \bibnamefont
			{McLean}}, \bibinfo {author} {\bibfnamefont {P.~K.}\ \bibnamefont {Patel}},
		\bibinfo {author} {\bibfnamefont {R.~B.}\ \bibnamefont {Stephens}},\ and\
		\bibinfo {author} {\bibfnamefont {F.~N.}\ \bibnamefont {Beg}},\ }\bibfield
	{title} {\bibinfo {title} {Focussing protons from a kilojoule laser for
			intense beam heating using proximal target structures},\ }\href
	{https://doi.org/10.1038/s41598-020-65554-4} {\bibfield  {journal} {\bibinfo
			{journal} {Sci. Rep.}\ }\textbf {\bibinfo {volume} {10}},\ \bibinfo {pages}
		{9415} (\bibinfo {year} {2020})}\BibitemShut {NoStop}%
	\bibitem [{\citenamefont {Kar}\ \emph {et~al.}(2008)\citenamefont {Kar},
		\citenamefont {Markey}, \citenamefont {Simpson}, \citenamefont {Bellei},
		\citenamefont {Green}, \citenamefont {Nagel}, \citenamefont {Kneip},
		\citenamefont {Carroll}, \citenamefont {Dromey}, \citenamefont {Willingale},
		\citenamefont {Clark}, \citenamefont {McKenna}, \citenamefont {Najmudin},
		\citenamefont {Krushelnick}, \citenamefont {Norreys}, \citenamefont {Clarke},
		\citenamefont {Neely}, \citenamefont {Borghesi},\ and\ \citenamefont
		{Zepf}}]{PhysRevLett.100.105004}%
	\BibitemOpen
	\bibfield  {author} {\bibinfo {author} {\bibfnamefont {S.}~\bibnamefont
			{Kar}}, \bibinfo {author} {\bibfnamefont {K.}~\bibnamefont {Markey}},
		\bibinfo {author} {\bibfnamefont {P.~T.}\ \bibnamefont {Simpson}}, \bibinfo
		{author} {\bibfnamefont {C.}~\bibnamefont {Bellei}}, \bibinfo {author}
		{\bibfnamefont {J.~S.}\ \bibnamefont {Green}}, \bibinfo {author}
		{\bibfnamefont {S.~R.}\ \bibnamefont {Nagel}}, \bibinfo {author}
		{\bibfnamefont {S.}~\bibnamefont {Kneip}}, \bibinfo {author} {\bibfnamefont
			{D.~C.}\ \bibnamefont {Carroll}}, \bibinfo {author} {\bibfnamefont
			{B.}~\bibnamefont {Dromey}}, \bibinfo {author} {\bibfnamefont
			{L.}~\bibnamefont {Willingale}}, \bibinfo {author} {\bibfnamefont {E.~L.}\
			\bibnamefont {Clark}}, \bibinfo {author} {\bibfnamefont {P.}~\bibnamefont
			{McKenna}}, \bibinfo {author} {\bibfnamefont {Z.}~\bibnamefont {Najmudin}},
		\bibinfo {author} {\bibfnamefont {K.}~\bibnamefont {Krushelnick}}, \bibinfo
		{author} {\bibfnamefont {P.}~\bibnamefont {Norreys}}, \bibinfo {author}
		{\bibfnamefont {R.~J.}\ \bibnamefont {Clarke}}, \bibinfo {author}
		{\bibfnamefont {D.}~\bibnamefont {Neely}}, \bibinfo {author} {\bibfnamefont
			{M.}~\bibnamefont {Borghesi}},\ and\ \bibinfo {author} {\bibfnamefont
			{M.}~\bibnamefont {Zepf}},\ }\bibfield  {title} {\bibinfo {title} {Dynamic
			control of laser-produced proton beams},\ }\href
	{https://doi.org/10.1103/PhysRevLett.100.105004} {\bibfield  {journal}
		{\bibinfo  {journal} {Phys. Rev. Lett.}\ }\textbf {\bibinfo {volume} {100}},\
		\bibinfo {pages} {105004} (\bibinfo {year} {2008})}\BibitemShut {NoStop}%
	\bibitem [{\citenamefont {Burza}\ \emph {et~al.}(2011)\citenamefont {Burza},
		\citenamefont {Gonoskov}, \citenamefont {Genoud}, \citenamefont {Persson},
		\citenamefont {Svensson}, \citenamefont {Quinn}, \citenamefont {McKenna},
		\citenamefont {Marklund},\ and\ \citenamefont {Wahlström}}]{Burza_2011}%
	\BibitemOpen
	\bibfield  {author} {\bibinfo {author} {\bibfnamefont {M.}~\bibnamefont
			{Burza}}, \bibinfo {author} {\bibfnamefont {A.}~\bibnamefont {Gonoskov}},
		\bibinfo {author} {\bibfnamefont {G.}~\bibnamefont {Genoud}}, \bibinfo
		{author} {\bibfnamefont {A.}~\bibnamefont {Persson}}, \bibinfo {author}
		{\bibfnamefont {K.}~\bibnamefont {Svensson}}, \bibinfo {author}
		{\bibfnamefont {M.}~\bibnamefont {Quinn}}, \bibinfo {author} {\bibfnamefont
			{P.}~\bibnamefont {McKenna}}, \bibinfo {author} {\bibfnamefont
			{M.}~\bibnamefont {Marklund}},\ and\ \bibinfo {author} {\bibfnamefont
			{C.-G.}\ \bibnamefont {Wahlström}},\ }\bibfield  {title} {\bibinfo {title}
		{Hollow microspheres as targets for staged laser-driven proton
			acceleration},\ }\href {https://doi.org/10.1088/1367-2630/13/1/013030}
	{\bibfield  {journal} {\bibinfo  {journal} {New J. Phys.}\ }\textbf {\bibinfo
			{volume} {13}},\ \bibinfo {pages} {013030} (\bibinfo {year}
		{2011})}\BibitemShut {NoStop}%
	\bibitem [{\citenamefont {Kar}\ \emph {et~al.}(2016{\natexlab{b}})\citenamefont
		{Kar}, \citenamefont {Ahmed}, \citenamefont {Prasad}, \citenamefont
		{Cerchez}, \citenamefont {Brauckmann}, \citenamefont {Aurand}, \citenamefont
		{Cantono}, \citenamefont {Hadjisolomou}, \citenamefont {Lewis}, \citenamefont
		{Macchi}, \citenamefont {Nersisyan}, \citenamefont {Robinson}, \citenamefont
		{Schroer}, \citenamefont {Swantusch}, \citenamefont {Zepf}, \citenamefont
		{Willi},\ and\ \citenamefont {Borghesi}}]{Kar2016}%
	\BibitemOpen
	\bibfield  {author} {\bibinfo {author} {\bibfnamefont {S.}~\bibnamefont
			{Kar}}, \bibinfo {author} {\bibfnamefont {H.}~\bibnamefont {Ahmed}}, \bibinfo
		{author} {\bibfnamefont {R.}~\bibnamefont {Prasad}}, \bibinfo {author}
		{\bibfnamefont {M.}~\bibnamefont {Cerchez}}, \bibinfo {author} {\bibfnamefont
			{S.}~\bibnamefont {Brauckmann}}, \bibinfo {author} {\bibfnamefont
			{B.}~\bibnamefont {Aurand}}, \bibinfo {author} {\bibfnamefont
			{G.}~\bibnamefont {Cantono}}, \bibinfo {author} {\bibfnamefont
			{P.}~\bibnamefont {Hadjisolomou}}, \bibinfo {author} {\bibfnamefont
			{C.~L.~S.}\ \bibnamefont {Lewis}}, \bibinfo {author} {\bibfnamefont
			{A.}~\bibnamefont {Macchi}}, \bibinfo {author} {\bibfnamefont
			{G.}~\bibnamefont {Nersisyan}}, \bibinfo {author} {\bibfnamefont {A.~P.~L.}\
			\bibnamefont {Robinson}}, \bibinfo {author} {\bibfnamefont {A.~M.}\
			\bibnamefont {Schroer}}, \bibinfo {author} {\bibfnamefont {M.}~\bibnamefont
			{Swantusch}}, \bibinfo {author} {\bibfnamefont {M.}~\bibnamefont {Zepf}},
		\bibinfo {author} {\bibfnamefont {O.}~\bibnamefont {Willi}},\ and\ \bibinfo
		{author} {\bibfnamefont {M.}~\bibnamefont {Borghesi}},\ }\bibfield  {title}
	{\bibinfo {title} {Guided post-acceleration of laser-driven ions by a
			miniature modular structure},\ }\href {https://doi.org/10.1038/ncomms10792}
	{\bibfield  {journal} {\bibinfo  {journal} {Nat. Commun.}\ }\textbf {\bibinfo
			{volume} {7}},\ \bibinfo {pages} {10792} (\bibinfo {year}
		{2016}{\natexlab{b}})}\BibitemShut {NoStop}%
	\bibitem [{\citenamefont {Schollmeier}\ \emph {et~al.}(2008)\citenamefont
		{Schollmeier}, \citenamefont {Becker}, \citenamefont {Gei\ss{}el},
		\citenamefont {Flippo}, \citenamefont {Bla\ifmmode \check{z}\else
			\v{z}\fi{}evi\ifmmode~\acute{c}\else \'{c}\fi{}}, \citenamefont {Gaillard},
		\citenamefont {Gautier}, \citenamefont {Gr\"uner}, \citenamefont {Harres},
		\citenamefont {Kimmel}, \citenamefont {N\"urnberg}, \citenamefont {Rambo},
		\citenamefont {Schramm}, \citenamefont {Schreiber}, \citenamefont
		{Sch\"utrumpf}, \citenamefont {Schwarz}, \citenamefont {Tahir}, \citenamefont
		{Atherton}, \citenamefont {Habs}, \citenamefont {Hegelich},\ and\
		\citenamefont {Roth}}]{PhysRevLett.101.055004}%
	\BibitemOpen
	\bibfield  {author} {\bibinfo {author} {\bibfnamefont {M.}~\bibnamefont
			{Schollmeier}}, \bibinfo {author} {\bibfnamefont {S.}~\bibnamefont {Becker}},
		\bibinfo {author} {\bibfnamefont {M.}~\bibnamefont {Gei\ss{}el}}, \bibinfo
		{author} {\bibfnamefont {K.~A.}\ \bibnamefont {Flippo}}, \bibinfo {author}
		{\bibfnamefont {A.}~\bibnamefont {Bla\ifmmode \check{z}\else
				\v{z}\fi{}evi\ifmmode~\acute{c}\else \'{c}\fi{}}}, \bibinfo {author}
		{\bibfnamefont {S.~A.}\ \bibnamefont {Gaillard}}, \bibinfo {author}
		{\bibfnamefont {D.~C.}\ \bibnamefont {Gautier}}, \bibinfo {author}
		{\bibfnamefont {F.}~\bibnamefont {Gr\"uner}}, \bibinfo {author}
		{\bibfnamefont {K.}~\bibnamefont {Harres}}, \bibinfo {author} {\bibfnamefont
			{M.}~\bibnamefont {Kimmel}}, \bibinfo {author} {\bibfnamefont
			{F.}~\bibnamefont {N\"urnberg}}, \bibinfo {author} {\bibfnamefont
			{P.}~\bibnamefont {Rambo}}, \bibinfo {author} {\bibfnamefont
			{U.}~\bibnamefont {Schramm}}, \bibinfo {author} {\bibfnamefont
			{J.}~\bibnamefont {Schreiber}}, \bibinfo {author} {\bibfnamefont
			{J.}~\bibnamefont {Sch\"utrumpf}}, \bibinfo {author} {\bibfnamefont
			{J.}~\bibnamefont {Schwarz}}, \bibinfo {author} {\bibfnamefont {N.~A.}\
			\bibnamefont {Tahir}}, \bibinfo {author} {\bibfnamefont {B.}~\bibnamefont
			{Atherton}}, \bibinfo {author} {\bibfnamefont {D.}~\bibnamefont {Habs}},
		\bibinfo {author} {\bibfnamefont {B.~M.}\ \bibnamefont {Hegelich}},\ and\
		\bibinfo {author} {\bibfnamefont {M.}~\bibnamefont {Roth}},\ }\bibfield
	{title} {\bibinfo {title} {Controlled transport and focusing of
			laser-accelerated protons with miniature magnetic devices},\ }\href
	{https://doi.org/10.1103/PhysRevLett.101.055004} {\bibfield  {journal}
		{\bibinfo  {journal} {Phys. Rev. Lett.}\ }\textbf {\bibinfo {volume} {101}},\
		\bibinfo {pages} {055004} (\bibinfo {year} {2008})}\BibitemShut {NoStop}%
	\bibitem [{\citenamefont {Nishiuchi}\ \emph {et~al.}(2009)\citenamefont
		{Nishiuchi}, \citenamefont {Daito}, \citenamefont {Ikegami}, \citenamefont
		{Daido}, \citenamefont {Mori}, \citenamefont {Orimo}, \citenamefont {Ogura},
		\citenamefont {Sagisaka}, \citenamefont {Yogo}, \citenamefont {Pirozhkov},
		\citenamefont {Sugiyama}, \citenamefont {Kiriyama}, \citenamefont {Okada},
		\citenamefont {Kanazawa}, \citenamefont {Kondo}, \citenamefont {Shimomura},
		\citenamefont {Tanoue}, \citenamefont {Nakai}, \citenamefont {Sasao},
		\citenamefont {Wakai}, \citenamefont {Sakaki}, \citenamefont {Bolton},
		\citenamefont {Choi}, \citenamefont {Sung}, \citenamefont {Lee},
		\citenamefont {Oishi}, \citenamefont {Fujii}, \citenamefont {Nemoto},
		\citenamefont {Souda}, \citenamefont {Noda}, \citenamefont {Iseki},\ and\
		\citenamefont {Yoshiyuki}}]{10.1063/1.3078291}%
	\BibitemOpen
	\bibfield  {author} {\bibinfo {author} {\bibfnamefont {M.}~\bibnamefont
			{Nishiuchi}}, \bibinfo {author} {\bibfnamefont {I.}~\bibnamefont {Daito}},
		\bibinfo {author} {\bibfnamefont {M.}~\bibnamefont {Ikegami}}, \bibinfo
		{author} {\bibfnamefont {H.}~\bibnamefont {Daido}}, \bibinfo {author}
		{\bibfnamefont {M.}~\bibnamefont {Mori}}, \bibinfo {author} {\bibfnamefont
			{S.}~\bibnamefont {Orimo}}, \bibinfo {author} {\bibfnamefont
			{K.}~\bibnamefont {Ogura}}, \bibinfo {author} {\bibfnamefont
			{A.}~\bibnamefont {Sagisaka}}, \bibinfo {author} {\bibfnamefont
			{A.}~\bibnamefont {Yogo}}, \bibinfo {author} {\bibfnamefont {A.~S.}\
			\bibnamefont {Pirozhkov}}, \bibinfo {author} {\bibfnamefont {H.}~\bibnamefont
			{Sugiyama}}, \bibinfo {author} {\bibfnamefont {H.}~\bibnamefont {Kiriyama}},
		\bibinfo {author} {\bibfnamefont {H.}~\bibnamefont {Okada}}, \bibinfo
		{author} {\bibfnamefont {S.}~\bibnamefont {Kanazawa}}, \bibinfo {author}
		{\bibfnamefont {S.}~\bibnamefont {Kondo}}, \bibinfo {author} {\bibfnamefont
			{T.}~\bibnamefont {Shimomura}}, \bibinfo {author} {\bibfnamefont
			{M.}~\bibnamefont {Tanoue}}, \bibinfo {author} {\bibfnamefont
			{Y.}~\bibnamefont {Nakai}}, \bibinfo {author} {\bibfnamefont
			{H.}~\bibnamefont {Sasao}}, \bibinfo {author} {\bibfnamefont
			{D.}~\bibnamefont {Wakai}}, \bibinfo {author} {\bibfnamefont
			{H.}~\bibnamefont {Sakaki}}, \bibinfo {author} {\bibfnamefont
			{P.}~\bibnamefont {Bolton}}, \bibinfo {author} {\bibfnamefont {I.~W.}\
			\bibnamefont {Choi}}, \bibinfo {author} {\bibfnamefont {J.~H.}\ \bibnamefont
			{Sung}}, \bibinfo {author} {\bibfnamefont {J.}~\bibnamefont {Lee}}, \bibinfo
		{author} {\bibfnamefont {Y.}~\bibnamefont {Oishi}}, \bibinfo {author}
		{\bibfnamefont {T.}~\bibnamefont {Fujii}}, \bibinfo {author} {\bibfnamefont
			{K.}~\bibnamefont {Nemoto}}, \bibinfo {author} {\bibfnamefont
			{H.}~\bibnamefont {Souda}}, \bibinfo {author} {\bibfnamefont
			{A.}~\bibnamefont {Noda}}, \bibinfo {author} {\bibfnamefont {Y.}~\bibnamefont
			{Iseki}},\ and\ \bibinfo {author} {\bibfnamefont {T.}~\bibnamefont
			{Yoshiyuki}},\ }\bibfield  {title} {\bibinfo {title} {Focusing and spectral
			enhancement of a repetition-rated, laser-driven, divergent multi-mev proton
			beam using permanent quadrupole magnets},\ }\href
	{https://doi.org/10.1063/1.3078291} {\bibfield  {journal} {\bibinfo
			{journal} {Appl. Phys. Lett.}\ }\textbf {\bibinfo {volume} {94}},\ \bibinfo
		{pages} {061107} (\bibinfo {year} {2009})}\BibitemShut {NoStop}%
	\bibitem [{\citenamefont {Ter-Avetisyan}\ \emph {et~al.}(2008)\citenamefont
		{Ter-Avetisyan}, \citenamefont {Schnürer}, \citenamefont {Polster},
		\citenamefont {Nickles},\ and\ \citenamefont
		{Sandner}}]{Ter-Avetisyan_Schnürer_Polster_Nickles_Sandner_2008}%
	\BibitemOpen
	\bibfield  {author} {\bibinfo {author} {\bibfnamefont {S.}~\bibnamefont
			{Ter-Avetisyan}}, \bibinfo {author} {\bibfnamefont {M.}~\bibnamefont
			{Schnürer}}, \bibinfo {author} {\bibfnamefont {R.}~\bibnamefont {Polster}},
		\bibinfo {author} {\bibfnamefont {P.}~\bibnamefont {Nickles}},\ and\ \bibinfo
		{author} {\bibfnamefont {W.}~\bibnamefont {Sandner}},\ }\bibfield  {title}
	{\bibinfo {title} {First demonstration of collimation and monochromatisation
			of a laser accelerated proton burst},\ }\href
	{https://doi.org/10.1017/S0263034608000712} {\bibfield  {journal} {\bibinfo
			{journal} {Laser and Particle Beams}\ }\textbf {\bibinfo {volume} {26}},\
		\bibinfo {pages} {637} (\bibinfo {year} {2008})}\BibitemShut {NoStop}%
	\bibitem [{\citenamefont {Wang}\ \emph {et~al.}(2010)\citenamefont {Wang},
		\citenamefont {Liu}, \citenamefont {Cai}, \citenamefont {Wang}, \citenamefont
		{Liu}, \citenamefont {Xia}, \citenamefont {Deng}, \citenamefont {Xu},
		\citenamefont {Leng}, \citenamefont {Li},\ and\ \citenamefont
		{Xu}}]{10.1063/1.3299363}%
	\BibitemOpen
	\bibfield  {author} {\bibinfo {author} {\bibfnamefont {W.}~\bibnamefont
			{Wang}}, \bibinfo {author} {\bibfnamefont {J.}~\bibnamefont {Liu}}, \bibinfo
		{author} {\bibfnamefont {Y.}~\bibnamefont {Cai}}, \bibinfo {author}
		{\bibfnamefont {C.}~\bibnamefont {Wang}}, \bibinfo {author} {\bibfnamefont
			{L.}~\bibnamefont {Liu}}, \bibinfo {author} {\bibfnamefont {C.}~\bibnamefont
			{Xia}}, \bibinfo {author} {\bibfnamefont {A.}~\bibnamefont {Deng}}, \bibinfo
		{author} {\bibfnamefont {Y.}~\bibnamefont {Xu}}, \bibinfo {author}
		{\bibfnamefont {Y.}~\bibnamefont {Leng}}, \bibinfo {author} {\bibfnamefont
			{R.}~\bibnamefont {Li}},\ and\ \bibinfo {author} {\bibfnamefont
			{Z.}~\bibnamefont {Xu}},\ }\bibfield  {title} {\bibinfo {title} {Angular and
			energy distribution of fast electrons emitted from a solid surface irradiated
			by femtosecond laser pulses in various conditions},\ }\href
	{https://doi.org/10.1063/1.3299363} {\bibfield  {journal} {\bibinfo
			{journal} {Phys. Plasmas}\ }\textbf {\bibinfo {volume} {17}},\ \bibinfo
		{pages} {023108} (\bibinfo {year} {2010})}\BibitemShut {NoStop}%
	\bibitem [{\citenamefont {Roth}\ \emph {et~al.}(2009)\citenamefont {Roth},
		\citenamefont {Alber}, \citenamefont {Bagnoud}, \citenamefont {Brown},
		\citenamefont {Clarke}, \citenamefont {Daido}, \citenamefont {Fernandez},
		\citenamefont {Flippo}, \citenamefont {Gaillard}, \citenamefont {Gauthier},
		\citenamefont {Geissel}, \citenamefont {Glenzer}, \citenamefont {Gregori},
		\citenamefont {Günther}, \citenamefont {Harres}, \citenamefont {Heathcote},
		\citenamefont {Kritcher}, \citenamefont {Kugland}, \citenamefont {LePape},
		\citenamefont {Li}, \citenamefont {Makita}, \citenamefont {Mithen},
		\citenamefont {Niemann}, \citenamefont {Nürnberg}, \citenamefont
		{Offermann}, \citenamefont {Otten}, \citenamefont {Pelka}, \citenamefont
		{Riley}, \citenamefont {Schaumann}, \citenamefont {Schollmeier},
		\citenamefont {Schütrumpf}, \citenamefont {Tampo}, \citenamefont
		{Tauschwitz},\ and\ \citenamefont {Tauschwitz}}]{Roth_2009}%
	\BibitemOpen
	\bibfield  {author} {\bibinfo {author} {\bibfnamefont {M.}~\bibnamefont
			{Roth}}, \bibinfo {author} {\bibfnamefont {I.}~\bibnamefont {Alber}},
		\bibinfo {author} {\bibfnamefont {V.}~\bibnamefont {Bagnoud}}, \bibinfo
		{author} {\bibfnamefont {C.~R.~D.}\ \bibnamefont {Brown}}, \bibinfo {author}
		{\bibfnamefont {R.}~\bibnamefont {Clarke}}, \bibinfo {author} {\bibfnamefont
			{H.}~\bibnamefont {Daido}}, \bibinfo {author} {\bibfnamefont
			{J.}~\bibnamefont {Fernandez}}, \bibinfo {author} {\bibfnamefont
			{K.}~\bibnamefont {Flippo}}, \bibinfo {author} {\bibfnamefont
			{S.}~\bibnamefont {Gaillard}}, \bibinfo {author} {\bibfnamefont
			{C.}~\bibnamefont {Gauthier}}, \bibinfo {author} {\bibfnamefont
			{M.}~\bibnamefont {Geissel}}, \bibinfo {author} {\bibfnamefont
			{S.}~\bibnamefont {Glenzer}}, \bibinfo {author} {\bibfnamefont
			{G.}~\bibnamefont {Gregori}}, \bibinfo {author} {\bibfnamefont
			{M.}~\bibnamefont {Günther}}, \bibinfo {author} {\bibfnamefont
			{K.}~\bibnamefont {Harres}}, \bibinfo {author} {\bibfnamefont
			{R.}~\bibnamefont {Heathcote}}, \bibinfo {author} {\bibfnamefont
			{A.}~\bibnamefont {Kritcher}}, \bibinfo {author} {\bibfnamefont
			{N.}~\bibnamefont {Kugland}}, \bibinfo {author} {\bibfnamefont
			{S.}~\bibnamefont {LePape}}, \bibinfo {author} {\bibfnamefont
			{B.}~\bibnamefont {Li}}, \bibinfo {author} {\bibfnamefont {M.}~\bibnamefont
			{Makita}}, \bibinfo {author} {\bibfnamefont {J.}~\bibnamefont {Mithen}},
		\bibinfo {author} {\bibfnamefont {C.}~\bibnamefont {Niemann}}, \bibinfo
		{author} {\bibfnamefont {F.}~\bibnamefont {Nürnberg}}, \bibinfo {author}
		{\bibfnamefont {D.}~\bibnamefont {Offermann}}, \bibinfo {author}
		{\bibfnamefont {A.}~\bibnamefont {Otten}}, \bibinfo {author} {\bibfnamefont
			{A.}~\bibnamefont {Pelka}}, \bibinfo {author} {\bibfnamefont
			{D.}~\bibnamefont {Riley}}, \bibinfo {author} {\bibfnamefont
			{G.}~\bibnamefont {Schaumann}}, \bibinfo {author} {\bibfnamefont
			{M.}~\bibnamefont {Schollmeier}}, \bibinfo {author} {\bibfnamefont
			{J.}~\bibnamefont {Schütrumpf}}, \bibinfo {author} {\bibfnamefont
			{M.}~\bibnamefont {Tampo}}, \bibinfo {author} {\bibfnamefont
			{A.}~\bibnamefont {Tauschwitz}},\ and\ \bibinfo {author} {\bibfnamefont
			{A.}~\bibnamefont {Tauschwitz}},\ }\bibfield  {title} {\bibinfo {title}
		{Proton acceleration experiments and warm dense matter research using high
			power lasers},\ }\href {https://doi.org/10.1088/0741-3335/51/12/124039}
	{\bibfield  {journal} {\bibinfo  {journal} {Plasma Phys. Control. Fusion}\
		}\textbf {\bibinfo {volume} {51}},\ \bibinfo {pages} {124039} (\bibinfo
		{year} {2009})}\BibitemShut {NoStop}%
	\bibitem [{\citenamefont {Ikegami}\ \emph {et~al.}(2009)\citenamefont
		{Ikegami}, \citenamefont {Nakamura}, \citenamefont {Iwashita}, \citenamefont
		{Shirai}, \citenamefont {Souda}, \citenamefont {Tajima}, \citenamefont
		{Tanabe}, \citenamefont {Tongu}, \citenamefont {Itoh}, \citenamefont
		{Shintaku}, \citenamefont {Yamazaki}, \citenamefont {Daido}, \citenamefont
		{Yogo}, \citenamefont {Orimo}, \citenamefont {Mori}, \citenamefont
		{Nishiuchi}, \citenamefont {Ogura}, \citenamefont {Sagisaka}, \citenamefont
		{Pirozhkov}, \citenamefont {Kiriyama}, \citenamefont {Kanazawa},
		\citenamefont {Kondo}, \citenamefont {Yamamoto}, \citenamefont {Shimomura},
		\citenamefont {Tanoue}, \citenamefont {Nakai}, \citenamefont {Akutsu},
		\citenamefont {Bulanov}, \citenamefont {Kimura}, \citenamefont {Oishi},
		\citenamefont {Nemoto}, \citenamefont {Tajima},\ and\ \citenamefont
		{Noda}}]{PhysRevSTAB.12.063501}%
	\BibitemOpen
	\bibfield  {author} {\bibinfo {author} {\bibfnamefont {M.}~\bibnamefont
			{Ikegami}}, \bibinfo {author} {\bibfnamefont {S.}~\bibnamefont {Nakamura}},
		\bibinfo {author} {\bibfnamefont {Y.}~\bibnamefont {Iwashita}}, \bibinfo
		{author} {\bibfnamefont {T.}~\bibnamefont {Shirai}}, \bibinfo {author}
		{\bibfnamefont {H.}~\bibnamefont {Souda}}, \bibinfo {author} {\bibfnamefont
			{Y.}~\bibnamefont {Tajima}}, \bibinfo {author} {\bibfnamefont
			{M.}~\bibnamefont {Tanabe}}, \bibinfo {author} {\bibfnamefont
			{H.}~\bibnamefont {Tongu}}, \bibinfo {author} {\bibfnamefont
			{H.}~\bibnamefont {Itoh}}, \bibinfo {author} {\bibfnamefont {H.}~\bibnamefont
			{Shintaku}}, \bibinfo {author} {\bibfnamefont {A.}~\bibnamefont {Yamazaki}},
		\bibinfo {author} {\bibfnamefont {H.}~\bibnamefont {Daido}}, \bibinfo
		{author} {\bibfnamefont {A.}~\bibnamefont {Yogo}}, \bibinfo {author}
		{\bibfnamefont {S.}~\bibnamefont {Orimo}}, \bibinfo {author} {\bibfnamefont
			{M.}~\bibnamefont {Mori}}, \bibinfo {author} {\bibfnamefont {M.}~\bibnamefont
			{Nishiuchi}}, \bibinfo {author} {\bibfnamefont {K.}~\bibnamefont {Ogura}},
		\bibinfo {author} {\bibfnamefont {A.}~\bibnamefont {Sagisaka}}, \bibinfo
		{author} {\bibfnamefont {A.~S.}\ \bibnamefont {Pirozhkov}}, \bibinfo {author}
		{\bibfnamefont {H.}~\bibnamefont {Kiriyama}}, \bibinfo {author}
		{\bibfnamefont {S.}~\bibnamefont {Kanazawa}}, \bibinfo {author}
		{\bibfnamefont {S.}~\bibnamefont {Kondo}}, \bibinfo {author} {\bibfnamefont
			{Y.}~\bibnamefont {Yamamoto}}, \bibinfo {author} {\bibfnamefont
			{T.}~\bibnamefont {Shimomura}}, \bibinfo {author} {\bibfnamefont
			{M.}~\bibnamefont {Tanoue}}, \bibinfo {author} {\bibfnamefont
			{Y.}~\bibnamefont {Nakai}}, \bibinfo {author} {\bibfnamefont
			{A.}~\bibnamefont {Akutsu}}, \bibinfo {author} {\bibfnamefont {S.~V.}\
			\bibnamefont {Bulanov}}, \bibinfo {author} {\bibfnamefont {T.}~\bibnamefont
			{Kimura}}, \bibinfo {author} {\bibfnamefont {Y.}~\bibnamefont {Oishi}},
		\bibinfo {author} {\bibfnamefont {K.}~\bibnamefont {Nemoto}}, \bibinfo
		{author} {\bibfnamefont {T.}~\bibnamefont {Tajima}},\ and\ \bibinfo {author}
		{\bibfnamefont {A.}~\bibnamefont {Noda}},\ }\bibfield  {title} {\bibinfo
		{title} {Radial focusing and energy compression of a laser-produced proton
			beam by a synchronous rf field},\ }\href
	{https://doi.org/10.1103/PhysRevSTAB.12.063501} {\bibfield  {journal}
		{\bibinfo  {journal} {Phys. Rev. ST Accel. Beams}\ }\textbf {\bibinfo
			{volume} {12}},\ \bibinfo {pages} {063501} (\bibinfo {year}
		{2009})}\BibitemShut {NoStop}%
	\bibitem [{\citenamefont {Toncian}\ \emph {et~al.}(2006)\citenamefont
		{Toncian}, \citenamefont {Borghesi}, \citenamefont {Fuchs}, \citenamefont
		{d'Humières}, \citenamefont {Antici}, \citenamefont {Audebert},
		\citenamefont {Brambrink}, \citenamefont {Cecchetti}, \citenamefont {Pipahl},
		\citenamefont {Romagnani},\ and\ \citenamefont
		{Willi}}]{doi:10.1126/science.1124412}%
	\BibitemOpen
	\bibfield  {author} {\bibinfo {author} {\bibfnamefont {T.}~\bibnamefont
			{Toncian}}, \bibinfo {author} {\bibfnamefont {M.}~\bibnamefont {Borghesi}},
		\bibinfo {author} {\bibfnamefont {J.}~\bibnamefont {Fuchs}}, \bibinfo
		{author} {\bibfnamefont {E.}~\bibnamefont {d'Humières}}, \bibinfo {author}
		{\bibfnamefont {P.}~\bibnamefont {Antici}}, \bibinfo {author} {\bibfnamefont
			{P.}~\bibnamefont {Audebert}}, \bibinfo {author} {\bibfnamefont
			{E.}~\bibnamefont {Brambrink}}, \bibinfo {author} {\bibfnamefont {C.~A.}\
			\bibnamefont {Cecchetti}}, \bibinfo {author} {\bibfnamefont {A.}~\bibnamefont
			{Pipahl}}, \bibinfo {author} {\bibfnamefont {L.}~\bibnamefont {Romagnani}},\
		and\ \bibinfo {author} {\bibfnamefont {O.}~\bibnamefont {Willi}},\ }\bibfield
	{title} {\bibinfo {title} {Ultrafast laser-driven microlens to focus and
			energy-select mega-electron volt protons},\ }\href
	{https://doi.org/10.1126/science.1124412} {\bibfield  {journal} {\bibinfo
			{journal} {Science}\ }\textbf {\bibinfo {volume} {312}},\ \bibinfo {pages}
		{410} (\bibinfo {year} {2006})}\BibitemShut {NoStop}%
	\bibitem [{\citenamefont {Ferguson}\ \emph {et~al.}(2023)\citenamefont
		{Ferguson}, \citenamefont {Martin}, \citenamefont {Ahmed}, \citenamefont
		{Aktan}, \citenamefont {Alanazi}, \citenamefont {Cerchez}, \citenamefont
		{Doria}, \citenamefont {Green}, \citenamefont {Greenwood}, \citenamefont
		{Odlozilik}, \citenamefont {Willi}, \citenamefont {Borghesi},\ and\
		\citenamefont {Kar}}]{Ferguson_2023}%
	\BibitemOpen
	\bibfield  {author} {\bibinfo {author} {\bibfnamefont {S.}~\bibnamefont
			{Ferguson}}, \bibinfo {author} {\bibfnamefont {P.}~\bibnamefont {Martin}},
		\bibinfo {author} {\bibfnamefont {H.}~\bibnamefont {Ahmed}}, \bibinfo
		{author} {\bibfnamefont {E.}~\bibnamefont {Aktan}}, \bibinfo {author}
		{\bibfnamefont {M.}~\bibnamefont {Alanazi}}, \bibinfo {author} {\bibfnamefont
			{M.}~\bibnamefont {Cerchez}}, \bibinfo {author} {\bibfnamefont
			{D.}~\bibnamefont {Doria}}, \bibinfo {author} {\bibfnamefont {J.~S.}\
			\bibnamefont {Green}}, \bibinfo {author} {\bibfnamefont {B.}~\bibnamefont
			{Greenwood}}, \bibinfo {author} {\bibfnamefont {B.}~\bibnamefont
			{Odlozilik}}, \bibinfo {author} {\bibfnamefont {O.}~\bibnamefont {Willi}},
		\bibinfo {author} {\bibfnamefont {M.}~\bibnamefont {Borghesi}},\ and\
		\bibinfo {author} {\bibfnamefont {S.}~\bibnamefont {Kar}},\ }\bibfield
	{title} {\bibinfo {title} {Dual stage approach to laser-driven helical coil
			proton acceleration},\ }\href {https://doi.org/10.1088/1367-2630/acaf99}
	{\bibfield  {journal} {\bibinfo  {journal} {New J. Phys.}\ }\textbf {\bibinfo
			{volume} {25}},\ \bibinfo {pages} {013006} (\bibinfo {year}
		{2023})}\BibitemShut {NoStop}%
	\bibitem [{\citenamefont {Offermann}\ \emph {et~al.}(2011)\citenamefont
		{Offermann}, \citenamefont {Flippo}, \citenamefont {Cobble}, \citenamefont
		{Schmitt}, \citenamefont {Gaillard}, \citenamefont {Bartal}, \citenamefont
		{Rose}, \citenamefont {Welch}, \citenamefont {Geissel},\ and\ \citenamefont
		{Schollmeier}}]{10.1063/1.3589476}%
	\BibitemOpen
	\bibfield  {author} {\bibinfo {author} {\bibfnamefont {D.~T.}\ \bibnamefont
			{Offermann}}, \bibinfo {author} {\bibfnamefont {K.~A.}\ \bibnamefont
			{Flippo}}, \bibinfo {author} {\bibfnamefont {J.}~\bibnamefont {Cobble}},
		\bibinfo {author} {\bibfnamefont {M.~J.}\ \bibnamefont {Schmitt}}, \bibinfo
		{author} {\bibfnamefont {S.~A.}\ \bibnamefont {Gaillard}}, \bibinfo {author}
		{\bibfnamefont {T.}~\bibnamefont {Bartal}}, \bibinfo {author} {\bibfnamefont
			{D.~V.}\ \bibnamefont {Rose}}, \bibinfo {author} {\bibfnamefont {D.~R.}\
			\bibnamefont {Welch}}, \bibinfo {author} {\bibfnamefont {M.}~\bibnamefont
			{Geissel}},\ and\ \bibinfo {author} {\bibfnamefont {M.}~\bibnamefont
			{Schollmeier}},\ }\bibfield  {title} {\bibinfo {title} {Characterization and
			focusing of light ion beams generated by ultra-intensely irradiated thin
			foils at the kilojoule scale a)},\ }\href {https://doi.org/10.1063/1.3589476}
	{\bibfield  {journal} {\bibinfo  {journal} {Phys. Plasmas}\ }\textbf
		{\bibinfo {volume} {18}},\ \bibinfo {pages} {056713} (\bibinfo {year}
		{2011})}\BibitemShut {NoStop}%
	\bibitem [{\citenamefont {Wang}\ \emph {et~al.}(2018)\citenamefont {Wang},
		\citenamefont {Shen}, \citenamefont {Zhang}, \citenamefont {Lu},
		\citenamefont {Li}, \citenamefont {Zhai}, \citenamefont {Li}, \citenamefont
		{Wang}, \citenamefont {Xu}, \citenamefont {Wang}, \citenamefont {Leng},
		\citenamefont {Liang}, \citenamefont {Li},\ and\ \citenamefont
		{Xu}}]{10.1063/1.5022347}%
	\BibitemOpen
	\bibfield  {author} {\bibinfo {author} {\bibfnamefont {W.~P.}\ \bibnamefont
			{Wang}}, \bibinfo {author} {\bibfnamefont {B.~F.}\ \bibnamefont {Shen}},
		\bibinfo {author} {\bibfnamefont {H.}~\bibnamefont {Zhang}}, \bibinfo
		{author} {\bibfnamefont {X.~M.}\ \bibnamefont {Lu}}, \bibinfo {author}
		{\bibfnamefont {J.~F.}\ \bibnamefont {Li}}, \bibinfo {author} {\bibfnamefont
			{S.~H.}\ \bibnamefont {Zhai}}, \bibinfo {author} {\bibfnamefont {S.~S.}\
			\bibnamefont {Li}}, \bibinfo {author} {\bibfnamefont {X.~L.}\ \bibnamefont
			{Wang}}, \bibinfo {author} {\bibfnamefont {R.~J.}\ \bibnamefont {Xu}},
		\bibinfo {author} {\bibfnamefont {C.}~\bibnamefont {Wang}}, \bibinfo {author}
		{\bibfnamefont {Y.~X.}\ \bibnamefont {Leng}}, \bibinfo {author}
		{\bibfnamefont {X.~Y.}\ \bibnamefont {Liang}}, \bibinfo {author}
		{\bibfnamefont {R.~X.}\ \bibnamefont {Li}},\ and\ \bibinfo {author}
		{\bibfnamefont {Z.~Z.}\ \bibnamefont {Xu}},\ }\bibfield  {title} {\bibinfo
		{title} {Multi-stage proton acceleration controlled by double beam image
			technique},\ }\href {https://doi.org/10.1063/1.5022347} {\bibfield  {journal}
		{\bibinfo  {journal} {Phys. Plasmas}\ }\textbf {\bibinfo {volume} {25}},\
		\bibinfo {pages} {063116} (\bibinfo {year} {2018})}\BibitemShut {NoStop}%
	\bibitem [{\citenamefont {Wang}\ \emph
		{et~al.}(2019{\natexlab{a}})\citenamefont {Wang}, \citenamefont {Shen},
		\citenamefont {Zhang}, \citenamefont {Lu}, \citenamefont {Li}, \citenamefont
		{Zhai}, \citenamefont {Li}, \citenamefont {Wang}, \citenamefont {Xu},
		\citenamefont {Wang}, \citenamefont {Leng}, \citenamefont {Liang},
		\citenamefont {Li},\ and\ \citenamefont {Xu}}]{10.1063/1.5088548}%
	\BibitemOpen
	\bibfield  {author} {\bibinfo {author} {\bibfnamefont {W.~P.}\ \bibnamefont
			{Wang}}, \bibinfo {author} {\bibfnamefont {B.~F.}\ \bibnamefont {Shen}},
		\bibinfo {author} {\bibfnamefont {H.}~\bibnamefont {Zhang}}, \bibinfo
		{author} {\bibfnamefont {X.~M.}\ \bibnamefont {Lu}}, \bibinfo {author}
		{\bibfnamefont {J.~F.}\ \bibnamefont {Li}}, \bibinfo {author} {\bibfnamefont
			{S.~H.}\ \bibnamefont {Zhai}}, \bibinfo {author} {\bibfnamefont {S.~S.}\
			\bibnamefont {Li}}, \bibinfo {author} {\bibfnamefont {X.~L.}\ \bibnamefont
			{Wang}}, \bibinfo {author} {\bibfnamefont {R.~J.}\ \bibnamefont {Xu}},
		\bibinfo {author} {\bibfnamefont {C.}~\bibnamefont {Wang}}, \bibinfo {author}
		{\bibfnamefont {Y.~X.}\ \bibnamefont {Leng}}, \bibinfo {author}
		{\bibfnamefont {X.~Y.}\ \bibnamefont {Liang}}, \bibinfo {author}
		{\bibfnamefont {R.~X.}\ \bibnamefont {Li}},\ and\ \bibinfo {author}
		{\bibfnamefont {Z.~Z.}\ \bibnamefont {Xu}},\ }\bibfield  {title} {\bibinfo
		{title} {Spectrum tailoring of low charge-to-mass ion beam by the
			triple-stage acceleration mechanism},\ }\href
	{https://doi.org/10.1063/1.5088548} {\bibfield  {journal} {\bibinfo
			{journal} {Phys. Plasmas}\ }\textbf {\bibinfo {volume} {26}},\ \bibinfo
		{pages} {043102} (\bibinfo {year} {2019}{\natexlab{a}})}\BibitemShut
	{NoStop}%
	\bibitem [{\citenamefont {Vieira}\ and\ \citenamefont
		{Mendon\ifmmode~\mbox{\c{c}}\else
			\c{c}\fi{}a}(2014)}]{PhysRevLett.112.215001}%
	\BibitemOpen
	\bibfield  {author} {\bibinfo {author} {\bibfnamefont {J.}~\bibnamefont
			{Vieira}}\ and\ \bibinfo {author} {\bibfnamefont {J.~T.}\ \bibnamefont
			{Mendon\ifmmode~\mbox{\c{c}}\else \c{c}\fi{}a}},\ }\bibfield  {title}
	{\bibinfo {title} {Nonlinear laser driven donut wakefields for positron and
			electron acceleration},\ }\href
	{https://doi.org/10.1103/PhysRevLett.112.215001} {\bibfield  {journal}
		{\bibinfo  {journal} {Phys. Rev. Lett.}\ }\textbf {\bibinfo {volume} {112}},\
		\bibinfo {pages} {215001} (\bibinfo {year} {2014})}\BibitemShut {NoStop}%
	\bibitem [{\citenamefont {Shi}\ \emph {et~al.}(2014)\citenamefont {Shi},
		\citenamefont {Shen}, \citenamefont {Zhang}, \citenamefont {Zhang},
		\citenamefont {Wang},\ and\ \citenamefont {Xu}}]{PhysRevLett.112.235001}%
	\BibitemOpen
	\bibfield  {author} {\bibinfo {author} {\bibfnamefont {Y.}~\bibnamefont
			{Shi}}, \bibinfo {author} {\bibfnamefont {B.}~\bibnamefont {Shen}}, \bibinfo
		{author} {\bibfnamefont {L.}~\bibnamefont {Zhang}}, \bibinfo {author}
		{\bibfnamefont {X.}~\bibnamefont {Zhang}}, \bibinfo {author} {\bibfnamefont
			{W.}~\bibnamefont {Wang}},\ and\ \bibinfo {author} {\bibfnamefont
			{Z.}~\bibnamefont {Xu}},\ }\bibfield  {title} {\bibinfo {title} {Light fan
			driven by a relativistic laser pulse},\ }\href
	{https://doi.org/10.1103/PhysRevLett.112.235001} {\bibfield  {journal}
		{\bibinfo  {journal} {Phys. Rev. Lett.}\ }\textbf {\bibinfo {volume} {112}},\
		\bibinfo {pages} {235001} (\bibinfo {year} {2014})}\BibitemShut {NoStop}%
	\bibitem [{\citenamefont {Wang}\ \emph
		{et~al.}(2015{\natexlab{a}})\citenamefont {Wang}, \citenamefont {Shen},
		\citenamefont {Zhang}, \citenamefont {Zhang}, \citenamefont {Shi},\ and\
		\citenamefont {Xu}}]{Wang2015}%
	\BibitemOpen
	\bibfield  {author} {\bibinfo {author} {\bibfnamefont {W.}~\bibnamefont
			{Wang}}, \bibinfo {author} {\bibfnamefont {B.}~\bibnamefont {Shen}}, \bibinfo
		{author} {\bibfnamefont {X.}~\bibnamefont {Zhang}}, \bibinfo {author}
		{\bibfnamefont {L.}~\bibnamefont {Zhang}}, \bibinfo {author} {\bibfnamefont
			{Y.}~\bibnamefont {Shi}},\ and\ \bibinfo {author} {\bibfnamefont
			{Z.}~\bibnamefont {Xu}},\ }\bibfield  {title} {\bibinfo {title} {Hollow
			screw-like drill in plasma using an intense laguerre-gaussian laser},\ }\href
	{https://doi.org/10.1038/srep08274} {\bibfield  {journal} {\bibinfo
			{journal} {Sci. Rep.}\ }\textbf {\bibinfo {volume} {5}},\ \bibinfo {pages}
		{8274} (\bibinfo {year} {2015}{\natexlab{a}})}\BibitemShut {NoStop}%
	\bibitem [{\citenamefont {Zhang}\ \emph {et~al.}(2015)\citenamefont {Zhang},
		\citenamefont {Shen}, \citenamefont {Shi}, \citenamefont {Wang},
		\citenamefont {Zhang}, \citenamefont {Wang}, \citenamefont {Xu},
		\citenamefont {Yi},\ and\ \citenamefont {Xu}}]{PhysRevLett.114.173901}%
	\BibitemOpen
	\bibfield  {author} {\bibinfo {author} {\bibfnamefont {X.}~\bibnamefont
			{Zhang}}, \bibinfo {author} {\bibfnamefont {B.}~\bibnamefont {Shen}},
		\bibinfo {author} {\bibfnamefont {Y.}~\bibnamefont {Shi}}, \bibinfo {author}
		{\bibfnamefont {X.}~\bibnamefont {Wang}}, \bibinfo {author} {\bibfnamefont
			{L.}~\bibnamefont {Zhang}}, \bibinfo {author} {\bibfnamefont
			{W.}~\bibnamefont {Wang}}, \bibinfo {author} {\bibfnamefont {J.}~\bibnamefont
			{Xu}}, \bibinfo {author} {\bibfnamefont {L.}~\bibnamefont {Yi}},\ and\
		\bibinfo {author} {\bibfnamefont {Z.}~\bibnamefont {Xu}},\ }\bibfield
	{title} {\bibinfo {title} {Generation of intense high-order vortex
			harmonics},\ }\href {https://doi.org/10.1103/PhysRevLett.114.173901}
	{\bibfield  {journal} {\bibinfo  {journal} {Phys. Rev. Lett.}\ }\textbf
		{\bibinfo {volume} {114}},\ \bibinfo {pages} {173901} (\bibinfo {year}
		{2015})}\BibitemShut {NoStop}%
	\bibitem [{\citenamefont {Zhu}\ \emph {et~al.}(2018)\citenamefont {Zhu},
		\citenamefont {Chen}, \citenamefont {Yu}, \citenamefont {Weng}, \citenamefont
		{Hu}, \citenamefont {McKenna},\ and\ \citenamefont
		{Sheng}}]{10.1063/1.5028555}%
	\BibitemOpen
	\bibfield  {author} {\bibinfo {author} {\bibfnamefont {X.-L.}\ \bibnamefont
			{Zhu}}, \bibinfo {author} {\bibfnamefont {M.}~\bibnamefont {Chen}}, \bibinfo
		{author} {\bibfnamefont {T.-P.}\ \bibnamefont {Yu}}, \bibinfo {author}
		{\bibfnamefont {S.-M.}\ \bibnamefont {Weng}}, \bibinfo {author}
		{\bibfnamefont {L.-X.}\ \bibnamefont {Hu}}, \bibinfo {author} {\bibfnamefont
			{P.}~\bibnamefont {McKenna}},\ and\ \bibinfo {author} {\bibfnamefont {Z.-M.}\
			\bibnamefont {Sheng}},\ }\bibfield  {title} {\bibinfo {title} {Bright
			attosecond $\gamma$-ray pulses from nonlinear compton scattering with
			laser-illuminated compound targets},\ }\href
	{https://doi.org/10.1063/1.5028555} {\bibfield  {journal} {\bibinfo
			{journal} {Appl. Phys. Lett.}\ }\textbf {\bibinfo {volume} {112}},\ \bibinfo
		{pages} {174102} (\bibinfo {year} {2018})}\BibitemShut {NoStop}%
	\bibitem [{\citenamefont {Ju}\ \emph {et~al.}(2018)\citenamefont {Ju},
		\citenamefont {Zhou}, \citenamefont {Jiang}, \citenamefont {Huang},
		\citenamefont {Zhang}, \citenamefont {Cai}, \citenamefont {Cao},
		\citenamefont {Qiao},\ and\ \citenamefont {Ruan}}]{Ju_2018}%
	\BibitemOpen
	\bibfield  {author} {\bibinfo {author} {\bibfnamefont {L.~B.}\ \bibnamefont
			{Ju}}, \bibinfo {author} {\bibfnamefont {C.~T.}\ \bibnamefont {Zhou}},
		\bibinfo {author} {\bibfnamefont {K.}~\bibnamefont {Jiang}}, \bibinfo
		{author} {\bibfnamefont {T.~W.}\ \bibnamefont {Huang}}, \bibinfo {author}
		{\bibfnamefont {H.}~\bibnamefont {Zhang}}, \bibinfo {author} {\bibfnamefont
			{T.~X.}\ \bibnamefont {Cai}}, \bibinfo {author} {\bibfnamefont {J.~M.}\
			\bibnamefont {Cao}}, \bibinfo {author} {\bibfnamefont {B.}~\bibnamefont
			{Qiao}},\ and\ \bibinfo {author} {\bibfnamefont {S.~C.}\ \bibnamefont
			{Ruan}},\ }\bibfield  {title} {\bibinfo {title} {Manipulating the topological
			structure of ultrarelativistic electron beams using laguerre-gaussian laser
			pulse},\ }\href {https://doi.org/10.1088/1367-2630/aac68a} {\bibfield
		{journal} {\bibinfo  {journal} {New J. Phys.}\ }\textbf {\bibinfo {volume}
			{20}},\ \bibinfo {pages} {063004} (\bibinfo {year} {2018})}\BibitemShut
	{NoStop}%
	\bibitem [{\citenamefont {Wang}\ \emph
		{et~al.}(2019{\natexlab{b}})\citenamefont {Wang}, \citenamefont {Jiang},
		\citenamefont {Shen}, \citenamefont {Yuan}, \citenamefont {Gan},
		\citenamefont {Zhang}, \citenamefont {Zhai},\ and\ \citenamefont
		{Xu}}]{PhysRevLett.122.024801}%
	\BibitemOpen
	\bibfield  {author} {\bibinfo {author} {\bibfnamefont {W.~P.}\ \bibnamefont
			{Wang}}, \bibinfo {author} {\bibfnamefont {C.}~\bibnamefont {Jiang}},
		\bibinfo {author} {\bibfnamefont {B.~F.}\ \bibnamefont {Shen}}, \bibinfo
		{author} {\bibfnamefont {F.}~\bibnamefont {Yuan}}, \bibinfo {author}
		{\bibfnamefont {Z.~M.}\ \bibnamefont {Gan}}, \bibinfo {author} {\bibfnamefont
			{H.}~\bibnamefont {Zhang}}, \bibinfo {author} {\bibfnamefont {S.~H.}\
			\bibnamefont {Zhai}},\ and\ \bibinfo {author} {\bibfnamefont {Z.~Z.}\
			\bibnamefont {Xu}},\ }\bibfield  {title} {\bibinfo {title} {New optical
			manipulation of relativistic vortex cutter},\ }\href
	{https://doi.org/10.1103/PhysRevLett.122.024801} {\bibfield  {journal}
		{\bibinfo  {journal} {Phys. Rev. Lett.}\ }\textbf {\bibinfo {volume} {122}},\
		\bibinfo {pages} {024801} (\bibinfo {year} {2019}{\natexlab{b}})}\BibitemShut
	{NoStop}%
	\bibitem [{\citenamefont {Pae}\ \emph {et~al.}(2020)\citenamefont {Pae},
		\citenamefont {Song}, \citenamefont {Ryu}, \citenamefont {Nam},\ and\
		\citenamefont {Kim}}]{Pae_2020}%
	\BibitemOpen
	\bibfield  {author} {\bibinfo {author} {\bibfnamefont {K.~H.}\ \bibnamefont
			{Pae}}, \bibinfo {author} {\bibfnamefont {H.}~\bibnamefont {Song}}, \bibinfo
		{author} {\bibfnamefont {C.-M.}\ \bibnamefont {Ryu}}, \bibinfo {author}
		{\bibfnamefont {C.~H.}\ \bibnamefont {Nam}},\ and\ \bibinfo {author}
		{\bibfnamefont {C.~M.}\ \bibnamefont {Kim}},\ }\bibfield  {title} {\bibinfo
		{title} {Low-divergence relativistic proton jet from a thin solid target
			driven by an ultra-intense circularly polarized laguerre-gaussian laser
			pulse},\ }\href {https://doi.org/10.1088/1361-6587/ab7d27} {\bibfield
		{journal} {\bibinfo  {journal} {Plasma Phys. Control. Fusion}\ }\textbf
		{\bibinfo {volume} {62}},\ \bibinfo {pages} {055009} (\bibinfo {year}
		{2020})}\BibitemShut {NoStop}%
	\bibitem [{\citenamefont {Blackman}\ \emph {et~al.}(2022)\citenamefont
		{Blackman}, \citenamefont {Shi}, \citenamefont {Klein}, \citenamefont
		{Cernaianu}, \citenamefont {Doria}, \citenamefont {Ghenuche},\ and\
		\citenamefont {Arefiev}}]{Blackman2022}%
	\BibitemOpen
	\bibfield  {author} {\bibinfo {author} {\bibfnamefont {D.~R.}\ \bibnamefont
			{Blackman}}, \bibinfo {author} {\bibfnamefont {Y.}~\bibnamefont {Shi}},
		\bibinfo {author} {\bibfnamefont {S.~R.}\ \bibnamefont {Klein}}, \bibinfo
		{author} {\bibfnamefont {M.}~\bibnamefont {Cernaianu}}, \bibinfo {author}
		{\bibfnamefont {D.}~\bibnamefont {Doria}}, \bibinfo {author} {\bibfnamefont
			{P.}~\bibnamefont {Ghenuche}},\ and\ \bibinfo {author} {\bibfnamefont
			{A.}~\bibnamefont {Arefiev}},\ }\bibfield  {title} {\bibinfo {title}
		{Electron acceleration from transparent targets irradiated by ultra-intense
			helical laser beams},\ }\href {https://doi.org/10.1038/s42005-022-00894-3}
	{\bibfield  {journal} {\bibinfo  {journal} {Commun. Phys.}\ }\textbf
		{\bibinfo {volume} {5}},\ \bibinfo {pages} {116} (\bibinfo {year}
		{2022})}\BibitemShut {NoStop}%
	\bibitem [{\citenamefont {Zhao}\ \emph {et~al.}(2022)\citenamefont {Zhao},
		\citenamefont {Hu}, \citenamefont {Lu}, \citenamefont {Zhang}, \citenamefont
		{Hu}, \citenamefont {Zhu}, \citenamefont {Sheng}, \citenamefont {Turcu},
		\citenamefont {Pukhov}, \citenamefont {Shao},\ and\ \citenamefont
		{Yu}}]{Zhao2022}%
	\BibitemOpen
	\bibfield  {author} {\bibinfo {author} {\bibfnamefont {J.}~\bibnamefont
			{Zhao}}, \bibinfo {author} {\bibfnamefont {Y.-T.}\ \bibnamefont {Hu}},
		\bibinfo {author} {\bibfnamefont {Y.}~\bibnamefont {Lu}}, \bibinfo {author}
		{\bibfnamefont {H.}~\bibnamefont {Zhang}}, \bibinfo {author} {\bibfnamefont
			{L.-X.}\ \bibnamefont {Hu}}, \bibinfo {author} {\bibfnamefont {X.-L.}\
			\bibnamefont {Zhu}}, \bibinfo {author} {\bibfnamefont {Z.-M.}\ \bibnamefont
			{Sheng}}, \bibinfo {author} {\bibfnamefont {I.~C.~E.}\ \bibnamefont {Turcu}},
		\bibinfo {author} {\bibfnamefont {A.}~\bibnamefont {Pukhov}}, \bibinfo
		{author} {\bibfnamefont {F.-Q.}\ \bibnamefont {Shao}},\ and\ \bibinfo
		{author} {\bibfnamefont {T.-P.}\ \bibnamefont {Yu}},\ }\bibfield  {title}
	{\bibinfo {title} {All-optical quasi-monoenergetic gev positron bunch
			generation by twisted laser fields},\ }\href
	{https://doi.org/10.1038/s42005-021-00797-9} {\bibfield  {journal} {\bibinfo
			{journal} {Commun. Phys.}\ }\textbf {\bibinfo {volume} {5}},\ \bibinfo
		{pages} {15} (\bibinfo {year} {2022})}\BibitemShut {NoStop}%
	\bibitem [{\citenamefont {Wang}\ \emph {et~al.}(2022)\citenamefont {Wang},
		\citenamefont {Dong}, \citenamefont {Shi}, \citenamefont {Leng},
		\citenamefont {Li},\ and\ \citenamefont {Xu}}]{10.1063/5.0121973}%
	\BibitemOpen
	\bibfield  {author} {\bibinfo {author} {\bibfnamefont {W.~P.}\ \bibnamefont
			{Wang}}, \bibinfo {author} {\bibfnamefont {H.}~\bibnamefont {Dong}}, \bibinfo
		{author} {\bibfnamefont {Z.~Y.}\ \bibnamefont {Shi}}, \bibinfo {author}
		{\bibfnamefont {Y.~X.}\ \bibnamefont {Leng}}, \bibinfo {author}
		{\bibfnamefont {R.~X.}\ \bibnamefont {Li}},\ and\ \bibinfo {author}
		{\bibfnamefont {Z.~Z.}\ \bibnamefont {Xu}},\ }\bibfield  {title} {\bibinfo
		{title} {Collimated particle acceleration by vortex laser-induced
			self-structured “plasma lens”},\ }\href
	{https://doi.org/10.1063/5.0121973} {\bibfield  {journal} {\bibinfo
			{journal} {Appl. Phys. Lett.}\ }\textbf {\bibinfo {volume} {121}},\ \bibinfo
		{pages} {214102} (\bibinfo {year} {2022})}\BibitemShut {NoStop}%
	\bibitem [{\citenamefont {Leblanc}\ \emph {et~al.}(2017)\citenamefont
		{Leblanc}, \citenamefont {Denoeud}, \citenamefont {Chopineau}, \citenamefont
		{Mennerat}, \citenamefont {Martin},\ and\ \citenamefont
		{Quéré}}]{Leblanc2017}%
	\BibitemOpen
	\bibfield  {author} {\bibinfo {author} {\bibfnamefont {A.}~\bibnamefont
			{Leblanc}}, \bibinfo {author} {\bibfnamefont {A.}~\bibnamefont {Denoeud}},
		\bibinfo {author} {\bibfnamefont {L.}~\bibnamefont {Chopineau}}, \bibinfo
		{author} {\bibfnamefont {G.}~\bibnamefont {Mennerat}}, \bibinfo {author}
		{\bibfnamefont {P.}~\bibnamefont {Martin}},\ and\ \bibinfo {author}
		{\bibfnamefont {F.}~\bibnamefont {Quéré}},\ }\bibfield  {title} {\bibinfo
		{title} {Plasma holograms for ultrahigh-intensity optics},\ }\href
	{https://doi.org/10.1038/nphys4007} {\bibfield  {journal} {\bibinfo
			{journal} {Nat. Phys.}\ }\textbf {\bibinfo {volume} {13}},\ \bibinfo {pages}
		{440} (\bibinfo {year} {2017})}\BibitemShut {NoStop}%
	\bibitem [{\citenamefont {Wang}\ \emph {et~al.}(2020)\citenamefont {Wang},
		\citenamefont {Jiang}, \citenamefont {Dong}, \citenamefont {Lu},
		\citenamefont {Li}, \citenamefont {Xu}, \citenamefont {Sun}, \citenamefont
		{Yu}, \citenamefont {Guo}, \citenamefont {Liang}, \citenamefont {Leng},
		\citenamefont {Li},\ and\ \citenamefont {Xu}}]{PhysRevLett.125.034801}%
	\BibitemOpen
	\bibfield  {author} {\bibinfo {author} {\bibfnamefont {W.~P.}\ \bibnamefont
			{Wang}}, \bibinfo {author} {\bibfnamefont {C.}~\bibnamefont {Jiang}},
		\bibinfo {author} {\bibfnamefont {H.}~\bibnamefont {Dong}}, \bibinfo {author}
		{\bibfnamefont {X.~M.}\ \bibnamefont {Lu}}, \bibinfo {author} {\bibfnamefont
			{J.~F.}\ \bibnamefont {Li}}, \bibinfo {author} {\bibfnamefont {R.~J.}\
			\bibnamefont {Xu}}, \bibinfo {author} {\bibfnamefont {Y.~J.}\ \bibnamefont
			{Sun}}, \bibinfo {author} {\bibfnamefont {L.~H.}\ \bibnamefont {Yu}},
		\bibinfo {author} {\bibfnamefont {Z.}~\bibnamefont {Guo}}, \bibinfo {author}
		{\bibfnamefont {X.~Y.}\ \bibnamefont {Liang}}, \bibinfo {author}
		{\bibfnamefont {Y.~X.}\ \bibnamefont {Leng}}, \bibinfo {author}
		{\bibfnamefont {R.~X.}\ \bibnamefont {Li}},\ and\ \bibinfo {author}
		{\bibfnamefont {Z.~Z.}\ \bibnamefont {Xu}},\ }\bibfield  {title} {\bibinfo
		{title} {Hollow plasma acceleration driven by a relativistic reflected hollow
			laser},\ }\href {https://doi.org/10.1103/PhysRevLett.125.034801} {\bibfield
		{journal} {\bibinfo  {journal} {Phys. Rev. Lett.}\ }\textbf {\bibinfo
			{volume} {125}},\ \bibinfo {pages} {034801} (\bibinfo {year}
		{2020})}\BibitemShut {NoStop}%
	\bibitem [{\citenamefont {Porat}\ \emph {et~al.}(2022)\citenamefont {Porat},
		\citenamefont {Lightman}, \citenamefont {Cohen},\ and\ \citenamefont
		{Pomerantz}}]{Porat_2022}%
	\BibitemOpen
	\bibfield  {author} {\bibinfo {author} {\bibfnamefont {E.}~\bibnamefont
			{Porat}}, \bibinfo {author} {\bibfnamefont {S.}~\bibnamefont {Lightman}},
		\bibinfo {author} {\bibfnamefont {I.}~\bibnamefont {Cohen}},\ and\ \bibinfo
		{author} {\bibfnamefont {I.}~\bibnamefont {Pomerantz}},\ }\bibfield  {title}
	{\bibinfo {title} {Spiral phase plasma mirror},\ }\href
	{https://doi.org/10.1088/2040-8986/ac79ba} {\bibfield  {journal} {\bibinfo
			{journal} {J. Opt.}\ }\textbf {\bibinfo {volume} {24}},\ \bibinfo {pages}
		{085501} (\bibinfo {year} {2022})}\BibitemShut {NoStop}%
	\bibitem [{\citenamefont {Wang}\ \emph
		{et~al.}(2015{\natexlab{b}})\citenamefont {Wang}, \citenamefont {Shen},
		\citenamefont {Zhang}, \citenamefont {Wang}, \citenamefont {Xu},
		\citenamefont {Yi},\ and\ \citenamefont {Shi}}]{10.1063/1.4917071}%
	\BibitemOpen
	\bibfield  {author} {\bibinfo {author} {\bibfnamefont {X.}~\bibnamefont
			{Wang}}, \bibinfo {author} {\bibfnamefont {B.}~\bibnamefont {Shen}}, \bibinfo
		{author} {\bibfnamefont {X.}~\bibnamefont {Zhang}}, \bibinfo {author}
		{\bibfnamefont {W.}~\bibnamefont {Wang}}, \bibinfo {author} {\bibfnamefont
			{J.}~\bibnamefont {Xu}}, \bibinfo {author} {\bibfnamefont {L.}~\bibnamefont
			{Yi}},\ and\ \bibinfo {author} {\bibfnamefont {Y.}~\bibnamefont {Shi}},\
	}\bibfield  {title} {\bibinfo {title} {High energy protons generation by two
			sequential laser pulses},\ }\href {https://doi.org/10.1063/1.4917071}
	{\bibfield  {journal} {\bibinfo  {journal} {Phys. Plasmas}\ }\textbf
		{\bibinfo {volume} {22}},\ \bibinfo {pages} {043106} (\bibinfo {year}
		{2015}{\natexlab{b}})}\BibitemShut {NoStop}%
	\bibitem [{\citenamefont {Brabetz}\ \emph {et~al.}(2015)\citenamefont
		{Brabetz}, \citenamefont {Busold}, \citenamefont {Cowan}, \citenamefont
		{Deppert}, \citenamefont {Jahn}, \citenamefont {Kester}, \citenamefont
		{Roth}, \citenamefont {Schumacher},\ and\ \citenamefont
		{Bagnoud}}]{10.1063/1.4905638}%
	\BibitemOpen
	\bibfield  {author} {\bibinfo {author} {\bibfnamefont {C.}~\bibnamefont
			{Brabetz}}, \bibinfo {author} {\bibfnamefont {S.}~\bibnamefont {Busold}},
		\bibinfo {author} {\bibfnamefont {T.}~\bibnamefont {Cowan}}, \bibinfo
		{author} {\bibfnamefont {O.}~\bibnamefont {Deppert}}, \bibinfo {author}
		{\bibfnamefont {D.}~\bibnamefont {Jahn}}, \bibinfo {author} {\bibfnamefont
			{O.}~\bibnamefont {Kester}}, \bibinfo {author} {\bibfnamefont
			{M.}~\bibnamefont {Roth}}, \bibinfo {author} {\bibfnamefont {D.}~\bibnamefont
			{Schumacher}},\ and\ \bibinfo {author} {\bibfnamefont {V.}~\bibnamefont
			{Bagnoud}},\ }\bibfield  {title} {\bibinfo {title} {Laser-driven ion
			acceleration with hollow laser beams},\ }\href
	{https://doi.org/10.1063/1.4905638} {\bibfield  {journal} {\bibinfo
			{journal} {Phys. Plasmas}\ }\textbf {\bibinfo {volume} {22}},\ \bibinfo
		{pages} {013105} (\bibinfo {year} {2015})}\BibitemShut {NoStop}%
	\bibitem [{\citenamefont {Longman}\ and\ \citenamefont
		{Fedosejevs}(2017)}]{Longman:17}%
	\BibitemOpen
	\bibfield  {author} {\bibinfo {author} {\bibfnamefont {A.}~\bibnamefont
			{Longman}}\ and\ \bibinfo {author} {\bibfnamefont {R.}~\bibnamefont
			{Fedosejevs}},\ }\bibfield  {title} {\bibinfo {title} {Mode conversion
			efficiency to laguerre-gaussian oam modes using spiral phase optics},\ }\href
	{https://doi.org/10.1364/OE.25.017382} {\bibfield  {journal} {\bibinfo
			{journal} {Opt. Express}\ }\textbf {\bibinfo {volume} {25}},\ \bibinfo
		{pages} {17382} (\bibinfo {year} {2017})}\BibitemShut {NoStop}%
	\bibitem [{\citenamefont {Ziegler}\ \emph {et~al.}(2024)\citenamefont
		{Ziegler}, \citenamefont {Göthel}, \citenamefont {Assenbaum}, \citenamefont
		{Bernert}, \citenamefont {Brack}, \citenamefont {Cowan}, \citenamefont
		{Dover}, \citenamefont {Gaus}, \citenamefont {Kluge}, \citenamefont {Kraft},
		\citenamefont {Kroll}, \citenamefont {Metzkes-Ng}, \citenamefont {Nishiuchi},
		\citenamefont {Prencipe}, \citenamefont {Püschel}, \citenamefont {Rehwald},
		\citenamefont {Reimold}, \citenamefont {Schlenvoigt}, \citenamefont
		{Umlandt}, \citenamefont {Vescovi}, \citenamefont {Schramm},\ and\
		\citenamefont {Zeil}}]{Ziegler2024}%
	\BibitemOpen
	\bibfield  {author} {\bibinfo {author} {\bibfnamefont {T.}~\bibnamefont
			{Ziegler}}, \bibinfo {author} {\bibfnamefont {I.}~\bibnamefont {Göthel}},
		\bibinfo {author} {\bibfnamefont {S.}~\bibnamefont {Assenbaum}}, \bibinfo
		{author} {\bibfnamefont {C.}~\bibnamefont {Bernert}}, \bibinfo {author}
		{\bibfnamefont {F.-E.}\ \bibnamefont {Brack}}, \bibinfo {author}
		{\bibfnamefont {T.~E.}\ \bibnamefont {Cowan}}, \bibinfo {author}
		{\bibfnamefont {N.~P.}\ \bibnamefont {Dover}}, \bibinfo {author}
		{\bibfnamefont {L.}~\bibnamefont {Gaus}}, \bibinfo {author} {\bibfnamefont
			{T.}~\bibnamefont {Kluge}}, \bibinfo {author} {\bibfnamefont
			{S.}~\bibnamefont {Kraft}}, \bibinfo {author} {\bibfnamefont
			{F.}~\bibnamefont {Kroll}}, \bibinfo {author} {\bibfnamefont
			{J.}~\bibnamefont {Metzkes-Ng}}, \bibinfo {author} {\bibfnamefont
			{M.}~\bibnamefont {Nishiuchi}}, \bibinfo {author} {\bibfnamefont
			{I.}~\bibnamefont {Prencipe}}, \bibinfo {author} {\bibfnamefont
			{T.}~\bibnamefont {Püschel}}, \bibinfo {author} {\bibfnamefont
			{M.}~\bibnamefont {Rehwald}}, \bibinfo {author} {\bibfnamefont
			{M.}~\bibnamefont {Reimold}}, \bibinfo {author} {\bibfnamefont {H.-P.}\
			\bibnamefont {Schlenvoigt}}, \bibinfo {author} {\bibfnamefont {M.~E.~P.}\
			\bibnamefont {Umlandt}}, \bibinfo {author} {\bibfnamefont {M.}~\bibnamefont
			{Vescovi}}, \bibinfo {author} {\bibfnamefont {U.}~\bibnamefont {Schramm}},\
		and\ \bibinfo {author} {\bibfnamefont {K.}~\bibnamefont {Zeil}},\ }\bibfield
	{title} {\bibinfo {title} {Laser-driven high-energy proton beams from
			cascaded acceleration regimes},\ }\href
	{https://doi.org/10.1038/s41567-024-02505-0} {\bibfield  {journal} {\bibinfo
			{journal} {Nat. Phys.}\ }\textbf {\bibinfo {volume} {20}},\ \bibinfo {pages}
		{1211} (\bibinfo {year} {2024})}\BibitemShut {NoStop}%
	\bibitem [{\citenamefont {Fryxell}\ \emph {et~al.}(2000)\citenamefont
		{Fryxell}, \citenamefont {Olson}, \citenamefont {Ricker}, \citenamefont
		{Timmes}, \citenamefont {Zingale}, \citenamefont {Lamb}, \citenamefont
		{MacNeice}, \citenamefont {Rosner}, \citenamefont {Truran},\ and\
		\citenamefont {Tufo}}]{Fryxell_2000}%
	\BibitemOpen
	\bibfield  {author} {\bibinfo {author} {\bibfnamefont {B.}~\bibnamefont
			{Fryxell}}, \bibinfo {author} {\bibfnamefont {K.}~\bibnamefont {Olson}},
		\bibinfo {author} {\bibfnamefont {P.}~\bibnamefont {Ricker}}, \bibinfo
		{author} {\bibfnamefont {F.~X.}\ \bibnamefont {Timmes}}, \bibinfo {author}
		{\bibfnamefont {M.}~\bibnamefont {Zingale}}, \bibinfo {author} {\bibfnamefont
			{D.~Q.}\ \bibnamefont {Lamb}}, \bibinfo {author} {\bibfnamefont
			{P.}~\bibnamefont {MacNeice}}, \bibinfo {author} {\bibfnamefont
			{R.}~\bibnamefont {Rosner}}, \bibinfo {author} {\bibfnamefont {J.~W.}\
			\bibnamefont {Truran}},\ and\ \bibinfo {author} {\bibfnamefont
			{H.}~\bibnamefont {Tufo}},\ }\bibfield  {title} {\bibinfo {title} {Flash: An
			adaptive mesh hydrodynamics code for modeling astrophysical thermonuclear
			flashes},\ }\href {https://doi.org/10.1086/317361} {\bibfield  {journal}
		{\bibinfo  {journal} {The Astrophysical Journal Supplement Series}\ }\textbf
		{\bibinfo {volume} {131}},\ \bibinfo {pages} {273} (\bibinfo {year}
		{2000})}\BibitemShut {NoStop}%
	\bibitem [{\citenamefont {Wang}\ \emph
		{et~al.}(2019{\natexlab{c}})\citenamefont {Wang}, \citenamefont {Jiang},
		\citenamefont {Li}, \citenamefont {Dong}, \citenamefont {Shen}, \citenamefont
		{Leng}, \citenamefont {Li},\ and\ \citenamefont
		{Xu}}]{Wang_Jiang_Li_Dong_Shen_Leng_Li_Xu_2019}%
	\BibitemOpen
	\bibfield  {author} {\bibinfo {author} {\bibfnamefont {W.}~\bibnamefont
			{Wang}}, \bibinfo {author} {\bibfnamefont {C.}~\bibnamefont {Jiang}},
		\bibinfo {author} {\bibfnamefont {S.}~\bibnamefont {Li}}, \bibinfo {author}
		{\bibfnamefont {H.}~\bibnamefont {Dong}}, \bibinfo {author} {\bibfnamefont
			{B.}~\bibnamefont {Shen}}, \bibinfo {author} {\bibfnamefont {Y.}~\bibnamefont
			{Leng}}, \bibinfo {author} {\bibfnamefont {R.}~\bibnamefont {Li}},\ and\
		\bibinfo {author} {\bibfnamefont {Z.}~\bibnamefont {Xu}},\ }\bibfield
	{title} {\bibinfo {title} {Monoenergetic proton beam accelerated by single
			reflection mechanism only during hole-boring stage},\ }\href
	{https://doi.org/10.1017/hpl.2019.40} {\bibfield  {journal} {\bibinfo
			{journal} {High Power Laser Sci. Eng.}\ }\textbf {\bibinfo {volume} {7}},\
		\bibinfo {pages} {e55} (\bibinfo {year} {2019}{\natexlab{c}})}\BibitemShut
	{NoStop}%
	\bibitem [{\citenamefont {Spitzer}\ and\ \citenamefont
		{H\"arm}(1953)}]{PhysRev.89.977}%
	\BibitemOpen
	\bibfield  {author} {\bibinfo {author} {\bibfnamefont {L.}~\bibnamefont
			{Spitzer}}\ and\ \bibinfo {author} {\bibfnamefont {R.}~\bibnamefont
			{H\"arm}},\ }\bibfield  {title} {\bibinfo {title} {Transport phenomena in a
			completely ionized gas},\ }\href {https://doi.org/10.1103/PhysRev.89.977}
	{\bibfield  {journal} {\bibinfo  {journal} {Phys. Rev.}\ }\textbf {\bibinfo
			{volume} {89}},\ \bibinfo {pages} {977} (\bibinfo {year} {1953})}\BibitemShut
	{NoStop}%
	\bibitem [{\citenamefont {Kemp}\ and\ \citenamefont {ter
			Vehn}(1998)}]{KEMP1998674}%
	\BibitemOpen
	\bibfield  {author} {\bibinfo {author} {\bibfnamefont {A.}~\bibnamefont
			{Kemp}}\ and\ \bibinfo {author} {\bibfnamefont {J.~M.}\ \bibnamefont {ter
				Vehn}},\ }\bibfield  {title} {\bibinfo {title} {An equation of state code for
			hot dense matter, based on the qeos description},\ }\href
	{https://doi.org/https://doi.org/10.1016/S0168-9002(98)00446-X} {\bibfield
		{journal} {\bibinfo  {journal} {Nuclear Instruments and Methods in Physics
				Research Section A: Accelerators, Spectrometers, Detectors and Associated
				Equipment}\ }\textbf {\bibinfo {volume} {415}},\ \bibinfo {pages} {674}
		(\bibinfo {year} {1998})}\BibitemShut {NoStop}%
	\bibitem [{\citenamefont {Eidmann}(1994)}]{Eidmann_1994}%
	\BibitemOpen
	\bibfield  {author} {\bibinfo {author} {\bibfnamefont {K.}~\bibnamefont
			{Eidmann}},\ }\bibfield  {title} {\bibinfo {title} {Radiation transport and
			atomic physics modeling in high-energy-density laser-produced plasmas},\
	}\href {https://doi.org/10.1017/S0263034600007709} {\bibfield  {journal}
		{\bibinfo  {journal} {Laser and Particle Beams}\ }\textbf {\bibinfo {volume}
			{12}},\ \bibinfo {pages} {223} (\bibinfo {year} {1994})}\BibitemShut
	{NoStop}%
	\bibitem [{\citenamefont {Arber}\ \emph {et~al.}(2015)\citenamefont {Arber},
		\citenamefont {Bennett}, \citenamefont {Brady}, \citenamefont
		{Lawrence-Douglas}, \citenamefont {Ramsay}, \citenamefont {Sircombe},
		\citenamefont {Gillies}, \citenamefont {Evans}, \citenamefont {Schmitz},
		\citenamefont {Bell},\ and\ \citenamefont {Ridgers}}]{Arber_2015}%
	\BibitemOpen
	\bibfield  {author} {\bibinfo {author} {\bibfnamefont {T.~D.}\ \bibnamefont
			{Arber}}, \bibinfo {author} {\bibfnamefont {K.}~\bibnamefont {Bennett}},
		\bibinfo {author} {\bibfnamefont {C.~S.}\ \bibnamefont {Brady}}, \bibinfo
		{author} {\bibfnamefont {A.}~\bibnamefont {Lawrence-Douglas}}, \bibinfo
		{author} {\bibfnamefont {M.~G.}\ \bibnamefont {Ramsay}}, \bibinfo {author}
		{\bibfnamefont {N.~J.}\ \bibnamefont {Sircombe}}, \bibinfo {author}
		{\bibfnamefont {P.}~\bibnamefont {Gillies}}, \bibinfo {author} {\bibfnamefont
			{R.~G.}\ \bibnamefont {Evans}}, \bibinfo {author} {\bibfnamefont
			{H.}~\bibnamefont {Schmitz}}, \bibinfo {author} {\bibfnamefont {A.~R.}\
			\bibnamefont {Bell}},\ and\ \bibinfo {author} {\bibfnamefont {C.~P.}\
			\bibnamefont {Ridgers}},\ }\bibfield  {title} {\bibinfo {title} {Contemporary
			particle-in-cell approach to laser-plasma modelling},\ }\href
	{https://doi.org/10.1088/0741-3335/57/11/113001} {\bibfield  {journal}
		{\bibinfo  {journal} {Plasma Phys. Control. Fusion}\ }\textbf {\bibinfo
			{volume} {57}},\ \bibinfo {pages} {113001} (\bibinfo {year}
		{2015})}\BibitemShut {NoStop}%
	\bibitem [{\citenamefont {Schollmeier}\ \emph {et~al.}(2014)\citenamefont
		{Schollmeier}, \citenamefont {Geissel}, \citenamefont {Sefkow},\ and\
		\citenamefont {Flippo}}]{10.1063/1.4870895}%
	\BibitemOpen
	\bibfield  {author} {\bibinfo {author} {\bibfnamefont {M.}~\bibnamefont
			{Schollmeier}}, \bibinfo {author} {\bibfnamefont {M.}~\bibnamefont
			{Geissel}}, \bibinfo {author} {\bibfnamefont {A.~B.}\ \bibnamefont
			{Sefkow}},\ and\ \bibinfo {author} {\bibfnamefont {K.~A.}\ \bibnamefont
			{Flippo}},\ }\bibfield  {title} {\bibinfo {title} {Improved spectral data
			unfolding for radiochromic film imaging spectroscopy of laser-accelerated
			proton beams},\ }\href {https://doi.org/10.1063/1.4870895} {\bibfield
		{journal} {\bibinfo  {journal} {Rev. Sci. Instrum.}\ }\textbf {\bibinfo
			{volume} {85}},\ \bibinfo {pages} {043305} (\bibinfo {year}
		{2014})}\BibitemShut {NoStop}%
\end{thebibliography}
%

\vspace{20pt}
\textbf{Acknowledgements:}
This study is supported by the Natural Science Foundation of China (Grant No. 12075306), Natural Science Foundation of Shanghai (Grant No. 22ZR1470900), Key Research Programs in Frontier Science (Grant No. ZDBSLY-SLH006), Strategic Priority Research Program of the Chinese Academy of Sciences (Grant No. XDA0380000), Shanghai Science and Technology Committee Program (Grant No. 22DZ1100300). We would like to thank C.-H. Hua, Wenjun Ma, and X.-Q. Yan for the analysis and calibration of the proton data in experiments.

\textbf{Author contributions:} 
R.-X. Li, Y.-X. Leng, and Z.-Z. Xu conceived the project. W.-P. Wang designed the experiments. W.-P. Wang, X.-Y. Sun, F.-Y. Sun performed the experiments and collected the data. X.-Y. Liang, Y. Xu, Z.-X. Zhang, F.-X. Wu, J.-B. Hu, J.-Y. Qian, and J.-C. Zhu constructed and ran the SULF laser system. X.-Y. Sun, F.-Y. Sun, Z.-X. Lv, Z.-Y. Shi conducted the simulations and analysed the experimental data. W.-P. Wang wrote the paper. All authors contributed to the experiments and discussions.

\textbf{Competing interests:}
The authors declare no competing interests.

\textbf{Data availability:}
The data that support the findings of this study are available from the corresponding authors on reasonable request.

\end{document}